\documentclass[sigconf]{acmart}

\newcommand{\myparatight}[1]{\smallskip\noindent{\bf {#1}:}~}

\usepackage{amsthm}
\usepackage{amsmath}

\usepackage{float}
\usepackage{graphics}
\usepackage{graphicx}
\usepackage{algorithm}
\usepackage{algorithmicx}
\usepackage{algpseudocode}
\usepackage{bm}
\usepackage{color}
\usepackage{multirow}
\usepackage{makecell}
\usepackage{balance}
\usepackage{subfig}

\usepackage{amssymb}

\allowdisplaybreaks

\newtheorem{assumption}{Assumption}
\newtheorem{thm}{Theorem}
\newtheorem{lem}{Lemma}
\newtheorem*{remark}{Remark}
\newtheorem{proposition}{Proposition}

\newcommand{\tabincell}[2]{\begin{tabular}{@{}#1@{}}#2\end{tabular}}

\algnewcommand\algorithmicforpara{\textbf{for}}
\algnewcommand\algorithmicdoinparallel{\textbf{do in parallel}}
\algdef{S}[FOR]{ForParallel}[1]{\algorithmicforpara\ #1\ \algorithmicdoinparallel}
\DeclareMathOperator*{\argmin}{arg\,min}

\copyrightyear{2022} 
\acmYear{2022} 
\setcopyright{acmcopyright}\acmConference[ACSAC '22]{Annual Computer Security Applications Conference}{December 5--9, 2022}{Austin, TX, USA}
\acmBooktitle{Annual Computer Security Applications Conference (ACSAC '22), December 5--9, 2022, Austin, TX, USA}
\acmPrice{15.00}
\acmDOI{10.1145/3564625.3567991}
\acmISBN{978-1-4503-9759-9/22/12}

\begin{document}

\title{AFLGuard: Byzantine-robust Asynchronous Federated Learning}

\author{Minghong Fang}
\affiliation{
	\institution{The Ohio State University and Duke University}
	\country{}
}

\author{Jia Liu}
\affiliation{
	\institution{The Ohio State University}
	\country{}
}

\author{Neil Zhenqiang Gong}
\affiliation{
	\institution{Duke University}
	\country{}
}

\author{Elizabeth S. Bentley}
\affiliation{
	\institution{Air Force Research Laboratory}
	\country{}
}

\begin{abstract}
	Federated learning (FL) is an emerging machine learning paradigm, in which clients jointly learn a model with the help of a cloud server. A fundamental challenge of  FL is that the clients are often heterogeneous, e.g., they have different computing powers, and thus the clients may send model updates to the server with substantially different delays. Asynchronous FL aims to address this challenge by enabling the server to update the model once any client's model update reaches it without waiting for other clients' model updates.  
	However, like synchronous FL, asynchronous FL is also vulnerable to poisoning attacks, in which malicious clients manipulate the model via poisoning their local data and/or model updates sent to the server.  Byzantine-robust FL aims to defend against poisoning attacks. In particular,  Byzantine-robust FL can learn an accurate model even if some clients are malicious and have Byzantine behaviors. However, most existing studies on  Byzantine-robust FL focused on synchronous FL, leaving asynchronous FL largely unexplored. In this work, we bridge this gap by proposing \emph{AFLGuard}, a Byzantine-robust asynchronous FL method. 
	We show that, both theoretically and empirically, AFLGuard is robust against various existing and adaptive poisoning attacks (both untargeted and targeted). Moreover, AFLGuard outperforms existing Byzantine-robust asynchronous FL methods. 
\end{abstract}

\ccsdesc[500]{Security and privacy~Systems security}

\keywords{Federated Learning, Poisoning Attacks, Byzantine Robustness}

\sloppy
\maketitle



\section{Introduction} \label{sec:intro}

\begin{figure*}[!t]
	\centering
	\subfloat[Synchronous FL]{\includegraphics[width=0.39 \textwidth]{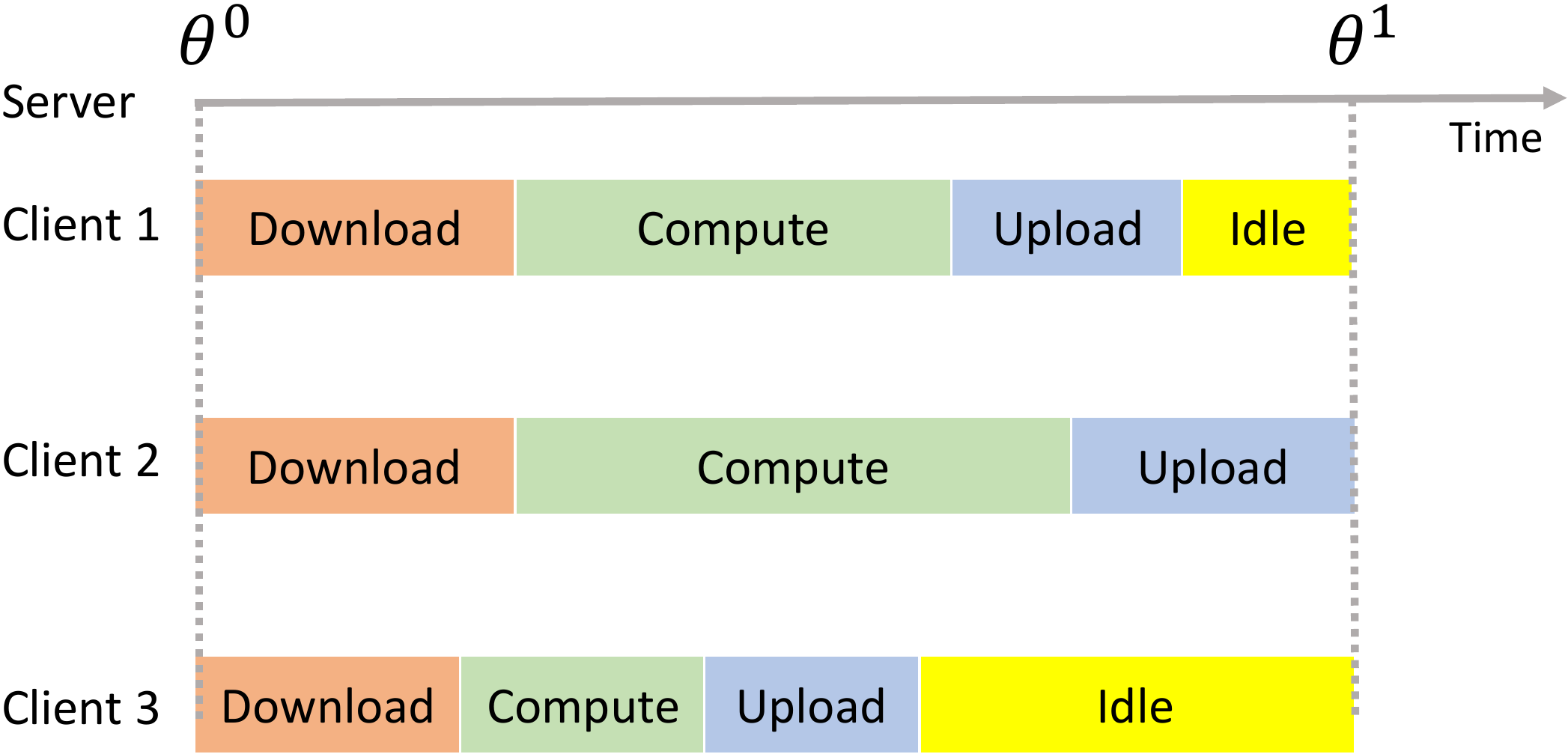}}
	\hskip4em
	\subfloat[Asynchronous FL]{\includegraphics[width=0.39 \textwidth]{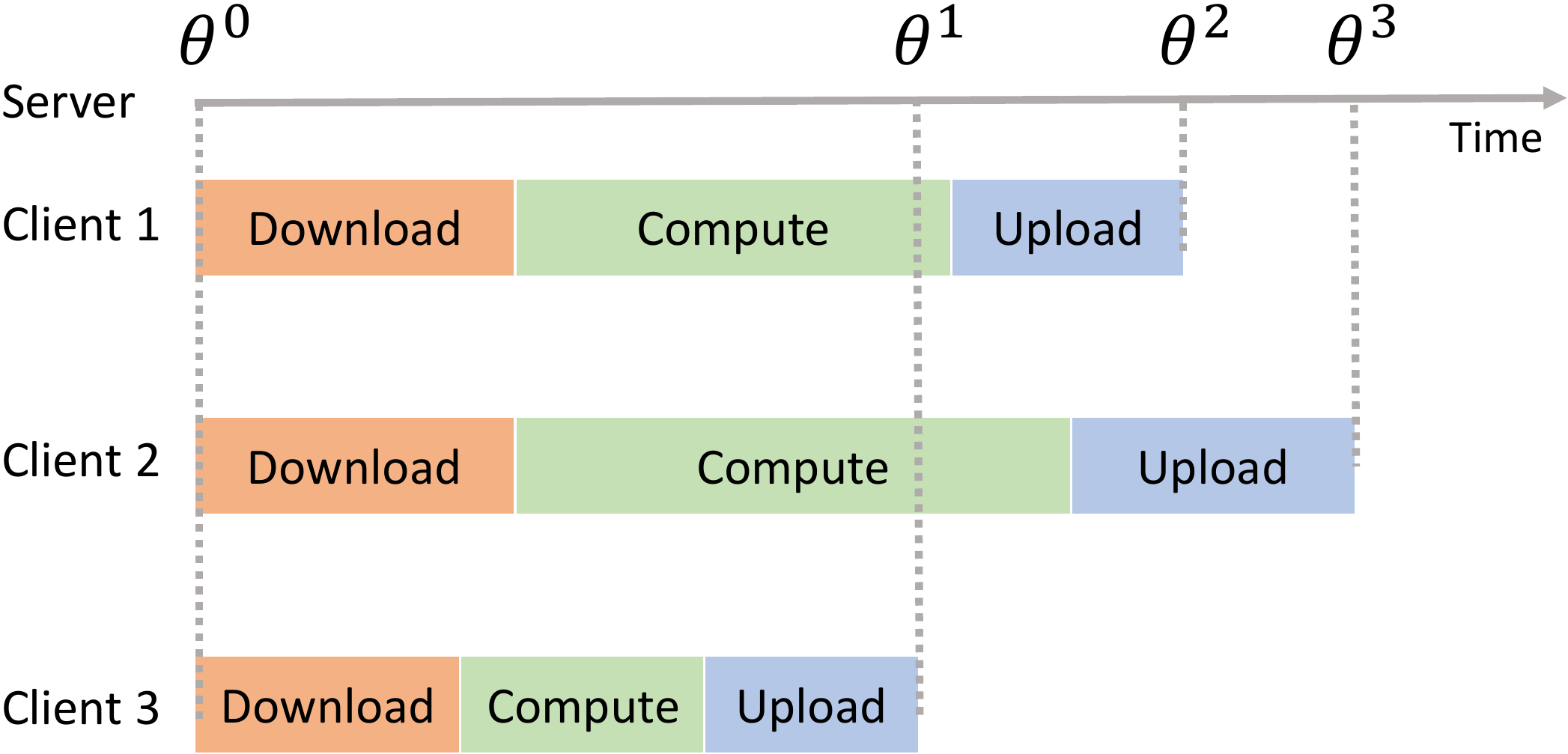}} 
	\vspace{.02in}
	\caption{Synchronous vs. asynchronous FL. ``Download'' means downloading the global model from the server. ``Compute'' means training a local model. ``Upload'' means sending the model update to the server. }
	\label{fig:syn_asy_fl_compare}
	\vspace{-0.02in}
\end{figure*}

\myparatight{Background and Motivation} Federated learning (FL)~\cite{kairouz2019advances, mcmahan2017communication} is an emerging distributed learning framework, which enables clients (e.g., smartphone, IoT device, edge device) to jointly train a \emph{global model} under the coordination of a cloud server. 
Specifically, the server maintains the global model and each client maintains a \emph{local model}. In each iteration, the server sends the current global model to the clients; a client trains its local model via fine-tuning  the global model using its local data, and the client sends the \emph{model update} (i.e., the difference between global model and local model) to the  server; and the  server aggregates the clients' model updates and uses them to update the global model.

Most existing FL methods are \emph{synchronous}~\cite{blanchard2017machine,mhamdi2018hidden,munoz2019byzantine,yin2018byzantine,cao2020fltrust,fung2020limitations}. Specifically, in each iteration of a synchronous FL, the server waits for the model updates from a large number of clients before aggregating them to update the global model. 
However, synchronous FL faces two key challenges. The first challenge is the so-called \emph{straggler problem}. Specifically, due to clients' unpredictable communication latency and/or heterogeneous  computing capabilities, some clients (i.e., stragglers) send their model updates to the server much later than others in each iteration, which substantially delays the update of the global model. Simply ignoring the stragglers' model updates would waste clients' computing resources and hurt accuracy of the global model~\cite{tandon2017gradient}. The second challenge is that synchronous FL is difficult to implement due to the high complexity in maintaining a perfectly synchronized global common clock.

 \emph{Asynchronous FL} aims to address the challenges of synchronous FL. Specifically, in asynchronous FL, the server updates the global model immediately upon receiving a client's model update without waiting for other clients' model updates. Due to the advantages of asynchronous FL, it has been widely incorporated in deep learning frameworks such as TensorFlow~\cite{abadi2016tensorflow} and PyTorch~\cite{paszke2019pytorch}, as well as deployed by industries, e.g., Meta~\cite{nguyen2022federated, huba2022papaya}. 
 However, like synchronous FL, asynchronous FL is also vulnerable to \emph{poisoning attacks}~\cite{damaskinos2018asynchronous,xie2020zeno++,yang2021basgd}, in which malicious clients  poison their local data and/or model updates to 
 guide the training process to converge to a bad global model. 
 Specifically, in \emph{untargeted poisoning attacks}~\cite{bhagoji2019analyzing,blanchard2017machine,fang2020local,shejwalkar2021manipulating}, the bad global model simply has a large error rate for indiscriminate testing inputs.  
 In \emph{targeted poisoning attacks}~\cite{chen2017targeted, shafahi2018poison, bagdasaryan2020backdoor,sun2019can}, the bad global model predicts attacker-chosen label for attacker-chosen testing inputs, but its predictions for other testing inputs are unaffected. 
 For instance, in \emph{backdoor attacks} (one type of targeted poisoning attacks)~\cite{bagdasaryan2020backdoor,sun2019can,wang2020attack,xie2019dba,nguyen2021flame}, the attacker-chosen testing inputs are inputs embedded with a backdoor trigger.

Byzantine-robust asynchronous FL aims to defend against poisoning attacks. However, it is highly challenging to design Byzantine-robust asynchronous FL.
To date, most existing Byzantine-robust FL methods (e.g., \cite{blanchard2017machine,mhamdi2018hidden,munoz2019byzantine,yin2018byzantine,cao2020fltrust,cao2019distributed}) are designed for synchronous FL.
Compared to synchronous FL, 
the key challenge in designing Byzantine-robust asynchronous FL stems from the fact that noisy model updates are inevitable.
Specifically, when a client sends its model update calculated based on a global model to the server, other clients may have already sent their model updates calculated based on the same global model to the server and thus the global model could have already been updated several times. As a result, delayed model updates are inevitably noisy with respect to the current global model. 
This asynchrony makes it difficult to distinguish between poisoned model updates from malicious clients  and the ``noisy'' model updates from benign clients.

\myparatight{Our Work} 
In this work,
 we propose {\em AFLGuard}, a Byzantine-robust asynchronous FL framework that addresses the aforementioned challenges. 
In AFLGuard, our key idea to handle the asynchrony complications is to equip the server with a small but clean training dataset, which we call \emph{trusted dataset}. The server (e.g., Meta, Google) can manually collect the trusted dataset for the learning task.  
 When receiving a model update from a client, the server computes a model update (called \emph{server model update}) based on its trusted dataset and the current global model.  
The server accepts the client's model update only if it does not deviate far from the server model update with respect to both direction and magnitude.  In particular, 
 if the magnitude of the difference vector between the client and server model updates is less than a certain fraction of the magnitude of the server model update, then the server uses the client's model update to update the global model. The updated global model is then sent to the client.

Interestingly, we show that this simple intuitive idea of AFLGuard enjoys strong theoretical guarantees. 
Specifically, under mild assumptions widely adopted by the Byzantine-robust FL community, we prove that the difference between the optimal global model parameters under no malicious clients and the global model parameters learnt by AFLGuard under an arbitrary number of malicious clients can be bounded. 
We also empirically evaluate AFLGuard and compare it with state-of-the-art Byzantine-robust asynchronous FL methods on a synthetic dataset and five real-world datasets.  
Our experimental results show that AFLGuard can defend against various existing and adaptive poisoning attacks when a large fraction of clients are malicious. 
Moreover, AFLGuard substantially outperforms existing Byzantine-robust asynchronous FL methods. 

We summarize our main contributions as follows:

\begin{list}{\labelitemi}{\leftmargin=2em \itemindent=-0.3em \itemsep=.2em}
	\item 
	
	We propose a  Byzantine-robust asynchronous FL  framework called AFLGuard to defend against poisoning attacks in asynchronous FL.
	
	\item 
	We theoretically show that AFLGuard is robust against an arbitrary number of malicious clients under mild assumptions commonly adopted by the Byzantine-robust FL community.
	
	\item 
	We conduct extensive experiments to  evaluate AFLGuard and compare it with state-of-the-art Byzantine-robust asynchronous FL methods on one synthetic and five real-world datasets. 

\end{list}



\section{Preliminaries and Related Work}

\subsection{Background on Federated Learning} 
\label{sec:background}

\myparatight{Notations} 
We use $\left\| \cdot \right\|$ to denote $\ell_2$-norm.
For any natural number $n$, we use $[n]$ to denote the set $\left\{1, 2,\cdots,n\right\}$.

\myparatight{Setup of Federated Learning (FL)} Suppose we have $n$ clients. Client $i$ has a local training dataset ${X}_i$, where $i=1,2,\cdots,n$. For simplicity, we denote by ${X}=\bigcup_{i=1}^n {X}_i$ the joint training dataset of the $n$ clients. An FL algorithm aims to solve an optimization problem, whose objective function is to find an optimal global model  $\bm{\theta}^*$ that minimizes the expected loss  $F(\bm{\theta})$ as follows:
\begin{align} \label{eqn_ERM}
\bm{\theta}^* = \argmin_{\bm{\theta} \in \Theta} F(\bm{\theta}) \triangleq \argmin_{\bm{\theta} \in \Theta} \mathbb{E}_{x\sim \mathcal{D}} \left[ f(\bm{\theta}, x) \right],
\end{align}
where $ \Theta \subseteq \mathbb{R}^d$ is the model-parameter space, $d$ is the dimension of the model-parameter space, $f$ is a loss function that evaluates the discrepancy between an output of a global model  and the corresponding ground truth,  the expectation $\mathbb{E}$ is taken with respect to the distribution of a training example $x$ (including both feature vector and label), and $\mathcal{D}$ is the training data distribution. In practice, the expectation is often approximated as the average loss of the training examples in the joint training dataset $X$, i.e., $\mathbb{E}_{x\sim \mathcal{D}} \left[ f(\bm{\theta}, x) \right] \approx \frac{1}{|X|}\sum_{x \in X} f(\bm{\theta}, x)$. 

In FL, the clients iteratively learn a global model with the coordination of a cloud server. In each iteration, synchronous FL waits for the information from multiple clients before using them to update the global model, while asynchronous FL updates the global model once the information from any client reaches it. Fig.~\ref{fig:syn_asy_fl_compare} illustrates the difference between synchronous FL and asynchronous FL~\cite{sys_asys_illu}. 

\myparatight{Synchronous FL} Synchronous FL performs three steps in each iteration. In the first step, the server sends the current global model to the clients or a selected subset of them. In the second step, a client trains its local model via fine-tuning the global model using its local training data, and it sends the model update (i.e., the difference between the global model and the local model) to the server. When all the selected clients have sent their model updates to the server, the server aggregates them and uses the aggregated model update to update the global model in the third step. For instance, in FedAvg~\cite{mcmahan2017communication}, the server computes the weighted average of the clients' model updates and uses it to update the global model in the third step.

\myparatight{Asynchronous FL} Synchronous FL requires the server to wait for the model updates from multiple clients before updating the global model, which is vulnerable to the straggler problem and delays the training process. In contrast, 
asynchronous FL updates the global model upon receiving a model update from any client~\cite{nguyen2022federated, huba2022papaya,chen2020asynchronous,xu2021asynchronous,xie2019asynchronous,chen2020vafl,van2020asynchronous}. Specifically, the server initializes the global model and sends it to all clients. Each client trains its local model via fine-tuning the global model based on its local training data, and sends the model update to the server. Upon receiving a model update, the server immediately updates the global model and sends the updated global model back to the client.

Formally, we denote by $\bm{\theta}^{t}$ the global model in the $t$th iteration. Moreover, we denote by $\bm{g}_i^{t}$ the model update from client $i$ that is calculated based on the global model $\bm{\theta}^{t}$. Suppose in the $t$th iteration, the server receives a model update  $\bm{g}_i^{t-\tau_i}$ from client $i$ that is calculated based on an earlier global model $\bm{\theta}^{t-\tau_i}$ in an earlier iteration $t-\tau_i$, where $\tau_i$ is the delay for the model update. The server updates the global model as follows:   
\begin{align}
\label{updaterule}
\bm{\theta}^{t+1} =  \bm{\theta}^{t} -\eta \bm{g}_i^{t-\tau_i},
\end{align}
where $\eta$ is the global learning rate. 

\emph{Asynchronous stochastic gradient descent (AsyncSGD)}~\cite{zheng2017asynchronous} is the most popular asynchronous FL method in non-adversarial settings. In AsyncSGD, a client simply fine-tunes the global model using one mini-batch of its local training data to obtain a local model. In other words, a client computes the gradient of the global model with respect to a random mini-batch of its local training data as the model update. Formally, $\bm{g}_i^{t} = \frac{1}{|B|}\sum_{x \in B} \triangledown f(\bm{\theta}^{t}, x)$, where $B$ is a mini-batch randomly sampled from $X_i$. Algorithm~\ref{AsynSGD_alg} shows AsyncSGD, where $T$ is the number of iterations. Note that for simplicity, we assume in all the compared FL methods and our AFLGuard, a client uses such gradient with respect to a random mini-batch of its local training data as the model update.

\begin{algorithm}[t]
	\caption{AsyncSGD.}
	\label{AsynSGD_alg}
	\begin{algorithmic}[1]
		 \Statex  \underline{Server:} 
		\State Initializes global model  $\bm{\theta}^{0} \in \Theta$ and sends it to all clients.
		\For {$t=0,1,2,\cdots,T-1$} 
		\State Upon receiving a model update $\bm{g}_i^{t-\tau_i}$ from client $i$, updates $\bm{\theta}^{t+1} =  \bm{\theta}^{t} -\eta \bm{g}_i^{t-\tau_i}$.
		\State Sends $\bm{\theta}^{t+1}$ to client $i$.
		\EndFor
		\Statex \underline{Client $i$, $i \in [n]$:} 
		\Repeat
			\State Receives a global model $\bm{\theta}^t$ from the server.
			\State Computes stochastic gradient $\bm{g}_i^t$ based on $\bm{\theta}^t$ and a random mini-batch of its local training data.
			\State Sends $\bm{g}_i^t$ to the server.
 		\Until{Convergence}
	\end{algorithmic} 
\end{algorithm}

\subsection{Byzantine-robust FL} \label{sec:related}

\myparatight{Poisoning Attacks to FL}
Poisoning attacks have been intensively studied in traditional ML systems, such as recommender systems~\cite{li2016data,fang2020influence,fang2018poisoning}, crowdsourcing systems~\cite{fang2021data,miao2018attack} and anomaly detectors~\cite{rubinstein2009antidote}.
Due to its distributed nature, FL is also vulnerable to poisoning attacks~\cite{bagdasaryan2020backdoor,bhagoji2019analyzing,fang2020local,cao2022mpaf}, in which malicious clients poison the global model via carefully manipulating their local training data and/or model updates. The malicious clients can be fake clients injected into the FL system by an attacker or genuine clients compromised by an attacker. 
Depending on the attack goal, poisoning attacks can be categorized into \emph{untargeted} and \emph{targeted}. In untargeted attacks, a poisoned global model has a large error rate for indiscriminate testing examples, leading to denial-of-service. In targeted attacks, a poisoned global model predicts attacker-chosen labels for attacker-chosen testing inputs, but its predictions for other testing inputs are unaffected. 

For instance, \emph{label flipping attack}~\cite{yin2018byzantine}, \emph{Gaussian attack}~\cite{blanchard2017machine}, and \emph{gradient deviation attack}~\cite{fang2020local} are examples of untargeted attacks. In particular, in the {label flipping attack}, the malicious clients flip the label $y$ of a local training example to $C-1-y$, where $C$ is the total number of labels and the labels are  $0,1,\cdots,C-1$. In the {Gaussian attack}, the malicious clients  draw their model updates from a Gaussian distribution with mean zero and a large standard deviation instead of computing them based on their local training data. In the {gradient deviation attack}, the model updates from the malicious clients are manipulated such that the global model update follows the reverse of the gradient direction (i.e., the direction where the global model should move without attacks). 

Backdoor attack~\cite{bagdasaryan2020backdoor,xie2019dba} is a popular targeted attack. For instance, in the backdoor attack in~\cite{bagdasaryan2020backdoor},  each malicious client replicates some of its local training examples; embeds a trigger (e.g., a patch on the right bottom corner of an image) into each replicated training input; and changes their labels to an attacker-chosen one. A malicious client calculates its model update  based on its original local training data and the replicated ones. Moreover, the malicious client scales up the model update by a scaling factor before sending it to the server. The poisoned global model would predict the attacker-chosen label for any input embedded with the same trigger, but the predictions for inputs without the trigger are not affected.

\myparatight{Byzantine-Robust Synchronous FL} Byzantine-robust FL aims to defend against poisoning attacks. Most existing Byzantine-robust FL methods focus on synchronous FL~\cite{blanchard2017machine,yin2018byzantine,cao2020fltrust}. Recall that a synchronous FL method has three steps in each iteration. These Byzantine-robust synchronous FL methods adopt robust aggregation rules in the third step. 
Roughly speaking, the key idea of a robust aggregation rule is to filter out ``outlier'' model updates before aggregating them to update the global model. For example, the Krum aggregation rule~\cite{blanchard2017machine} outputs the model update with the minimal sum of distances to its $n -m -2$ neighbors, where $n$ and $m$ are the numbers of total and malicious clients, respectively.
Since these methods are designed to aggregate model updates from multiple clients, they are not applicable to asynchronous FL, which updates the global model using one model update. Other defenses for synchronous FL include provably secure defenses to prevent poisoning attacks~\cite{cao2020provably} and methods to detect malicious clients~\cite{zhang2022fldetector}.

\myparatight{Byzantine-Robust Asynchronous FL}
To the best of our knowledge, the works most related to ours are~\cite{damaskinos2018asynchronous,xie2020zeno++,yang2021basgd}.
Specifically, Kardam~\cite{damaskinos2018asynchronous} maintains a Lipschitz coefficient for each client based on its latest model update sent to the server. 
The server uses a model update from a client to update the global model only if its Lipschitz coefficient is smaller than the median Lipschitz coefficient of all clients. 
 BASGD~\cite{yang2021basgd} is a non-conventional asynchronous FL method that uses multiple clients' model updates to update the global model. Specifically, the server holds several buffers and maps each client's model update into one of them.  When all buffers are non-empty, the server  computes the average of the model updates in each buffer, takes the median or trimmed-mean of the average model updates, and uses it to update the global model. 
In Zeno++~\cite{xie2020zeno++}, the server filters clients' model updates based on a trusted dataset. 
The server computes a server model update based on the trusted dataset.
After receiving a model update from any client, the server computes the cosine similarity between the client model update and server model update.
If the cosine similarity is positive, then the server normalizes the client model update.
Note that FLTrust~\cite{cao2020fltrust}, a synchronous FL method,
uses the similar technique as in Zeno++ to filter out malicious information.

\myparatight{Differences between AFLGuard and Zeno++}
Both our AFLGuard and Zeno++ use a trusted dataset on the server. However, they use it in different ways. 
Zeno++ simply treats a client’s model update as benign if it is positively correlated with the server model update. Due to delays on both client and server sides and the distribution shift between the trusted and clients' training data, the server’s and benign clients’ model updates may not be positively correlated. In AFLGuard, a client’s model update is considered benign if it does not deviate substantially from the server’s model update in both direction and magnitude.




\section{Problem Formulation}

\myparatight{Threat Model}
The attacker controls some malicious clients, which could be  genuine clients compromised by the attacker or fake clients injected by the attacker. 
The attacker does not compromise the server. 
The malicious clients could send arbitrary model updates to the server.
The attacker could have different degree of knowledge about the FL system~\cite{fang2020local,cao2020fltrust}, i.e., \emph{partial knowledge} and \emph{full knowledge}.
In the partial-knowledge setting, the attacker knows the local training data and model updates on the malicious clients.
In the full-knowledge scenario, the attacker has full knowledge of the FL system.
In particular, the attacker knows the local training data and model updates on
all clients, as well as the FL method and its parameter settings.
Note that the attacker in the full-knowledge setting is much stronger than that of partial-knowledge setting~\cite{fang2020local}.
Following~\cite{cao2020fltrust}, we use the full-knowledge attack setting to evaluate the security of our defense in the worst case. In other words, our defense is more secure against weaker attacks.

\myparatight{Defense Goals}
We aim to design an asynchronous FL method that achieves the following two goals: i) the method should be as accurate as AsyncSGD in \emph{non-adversarial settings}. In other words, when all clients are benign, our method  should learn as an accurate global model as AsyncSGD; and ii) the method should be robust against both existing and adaptive poisoning attacks in adversarial settings. Adaptive poisoning attacks refer to attacks that are tailored to the proposed method.

\myparatight{Server’s Capability and Knowledge}
We assume the server holds a small clean dataset, which we call \emph{trusted dataset}. This assumption is reasonable in practice  because it is quite affordable for a service provider to collect and  verify a small trusted dataset for the learning task. 
For instance, Google uses FL for the next word prediction in a virtual keyboard application called Gboard~\cite{gboard}; and Google can collect a trusted dataset from its employees. The trusted dataset does not need to follow the same distribution as the joint training dataset $X$. As our experimental results will show, once the trusted dataset distribution does not deviate substantially from the joint training data distribution, our method is effective. We acknowledge that the trusted dataset should be clean, and our method may not be robust when the trusted dataset is poisoned.

\section{AFLGuard}
\label{sec:method}

 \begin{figure}[!t]
	\centering
	{\includegraphics[width= 0.42\textwidth]{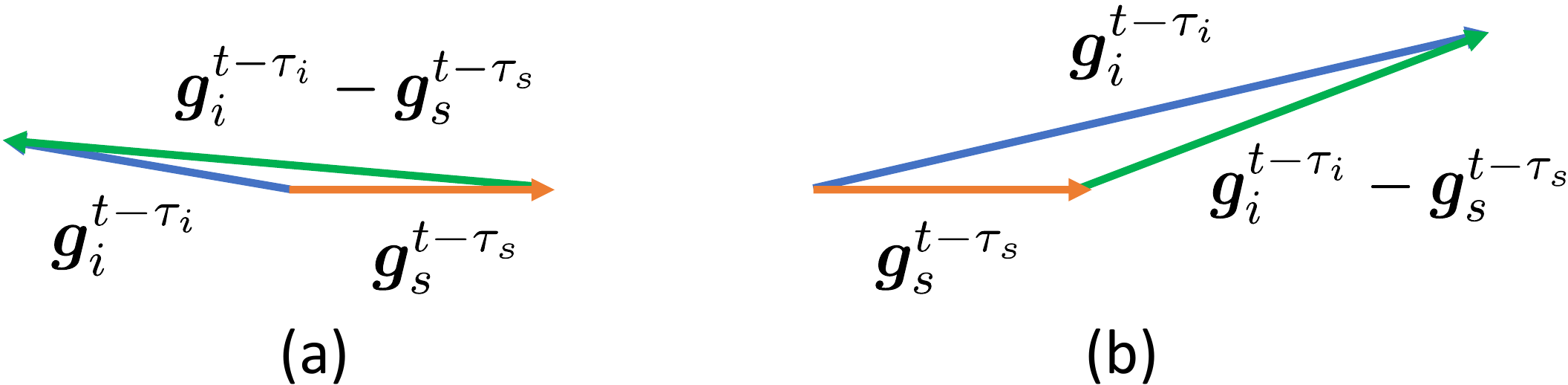}}
	\vspace{.05in}
	\caption{Illustration of our acceptance criteria.  $\bm{g}_i^{t-\tau_i}$ and $ \bm{g}_s^{t-\tau_s}$ are client model update and server model update, respectively.
		(a) the direction of $\bm{g}_i^{t-\tau_i}$ deviates substantially from that of $\bm{g}_s^{t-\tau_s}$. (b) the magnitude of  $\bm{g}_i^{t-\tau_i}$ deviates substantially from that of  $\bm{g}_s^{t-\tau_s}$. The server rejects the client model update in both cases.}
	\label{fig:afl_guard}
	\vspace{-.1in}
\end{figure}

\myparatight{Intuitions} The key of our AFLGuard is a criteria to decide whether the server should accept a client's model update to update the global model or not. Ideally, if a model update is from a malicious client performing poisoning attacks, then the server should not use it to update the global model. Our key observation is that, in poisoning attacks, malicious clients often manipulate the directions and/or the magnitudes of their model updates. Therefore, we consider both the direction and magnitude of a client's model update when deciding whether it should be accepted to update the global model or not. Specifically, the server computes a model update (called \emph{server model update}) based on its own trusted dataset. When a client's model update deviates substantially from the server model update with respect to  direction and/or magnitude, it is rejected. 

\myparatight{Acceptance Criteria} Suppose in the $t$th iteration, the server receives a model update $\bm{g}_i^{t-\tau_i}$ from a client $i \in [n]$, where $\tau_i$ is the delay.  Client $i$ calculated the model update $\bm{g}_i^{t-\tau_i}$ based on the global model $\bm{\theta}^{t-\tau_i}$, i.e., the server previously sent the global model $\bm{\theta}^{t-\tau_i}$ to client $i$ in the $(t-\tau_i)$th iteration. 
Moreover, in the $t$th iteration, the server has a model update  $\bm{g}_s^{t-\tau_s}$ based on its trusted dataset, where $\tau_s$ is the delay (called  \textit{server delay}) for the server model update. Specifically, the server trains a local model via fine-tuning the global model $\bm{\theta}^{t-\tau_s}$ using its trusted dataset, and the model update $\bm{g}_s^{t-\tau_s}$ is the difference between the global model $\bm{\theta}^{t-\tau_s}$ and the local model. We note that we assume the server model update can have a delay $\tau_s$, i.e., the server is not required to compute the model update using the global model $\bm{\theta}^{t}$ in the $t$th iteration. Instead, the server can compute a model update in every $\tau_s$ iterations.

The server accepts $\bm{g}_i^{t-\tau_i}$ if i) the direction of  $\bm{g}_i^{t-\tau_i}$ does not deviate dramatically from that of $\bm{g}_s^{t-\tau_s}$ and ii) the magnitude of $\bm{g}_i^{t-\tau_i}$ is similar to that of $\bm{g}_s^{t-\tau_s}$. 
Formally,   
the server accepts $\bm{g}_i^{t-\tau_i}$ if the following inequality is satisfied:
\begin{align} \label{guard_lip_sel_eq}
\left\|\bm{g}_i^{t-\tau_i} -  \bm{g}_s^{t-\tau_s}  \right\| \le \lambda \left\| \bm{g}_s^{t-\tau_s} \right\|,
\end{align}
where the parameter $\lambda>0$ can be viewed as a control knob: if $\lambda$ is too small, the server could potentially reject some model updates from benign clients; on the other hand, if $\lambda$ is  too large, the server could falsely accept some model updates from malicious clients. Fig.~\ref{fig:afl_guard} illustrates our acceptance criteria. Once a client's model update is accepted, the server uses it to update the global model based on Eq.~(\ref{updaterule}).

\myparatight{Algorithm of AFLGuard}
We summarize our AFLGuard algorithm in Algorithm~\ref{AFLGuard_lip_alg}. Note that Algorithm~\ref{AFLGuard_lip_alg} only shows the learning procedure of AFLGuard on the server side.
The learning procedure on the client side is the same as that of Algorithm~\ref{AsynSGD_alg} and thus we omit it  for brevity. In the $t$th iteration, the server decides whether to accept a client's model update or not based on Eq.~(\ref{guard_lip_sel_eq}). If yes, the server uses it to update the global model and sends the updated global model back to the client. Otherwise the server does not update the global model and sends the current global model back to the client.

\begin{algorithm}[t]
	\caption{Our AFLGuard.}
	\label{AFLGuard_lip_alg}
	\begin{algorithmic}[1]
		\Statex  \underline{Server:} 
		\State Initializes global model $\bm{\theta}^{0} \in \Theta$ and sends it to all clients.
		\For {$t=0,1,2,\cdots,T-1$} 
		\State Upon receiving a model update $\bm{g}_i^{t-\tau_i}$ from a client $i$, retrieves the server model update $\bm{g}_s^{t-\tau_s}$. 
		\If  {$\left\|\bm{g}_i^{t-\tau_i} -\bm{g}_s^{t-\tau_s} \right\| \le \lambda \left\|\bm{g}_s^{t-\tau_s} \right\| $} 
		\State Updates the global model $\bm{\theta}^{t+1} =  \bm{\theta}^{t} -\eta \bm{g}_i^{t-\tau_i}$.
    	\Else
    	\State Does not update the global model, i.e., $\bm{\theta}^{t+1} =  \bm{\theta}^{t}$.

		\EndIf
		\State Sends the global model  $\bm{\theta}^{t+1}$ to client $i$.

		\EndFor
		
	\end{algorithmic} 
\end{algorithm}

\section{Theoretical  Security Analysis}
\label{sec:sec_ana}

We theoretically analyze the security/robustness of AFLGuard. In particular, we show that the difference between the optimal global model $\bm{\theta}^*$ under no malicious clients and the global model learnt by AFLGuard with malicious clients can be bounded under some assumptions. We note that simple models like regression can satisfy these assumptions, while more complex models like neural networks may not. Therefore, in the next section, we will empirically evaluate our method on complex neural networks.

For convenience, we define $\bm{V}$ as the $d$-dimensional unit vector space $\bm{V} \stackrel{\text{def}} = \{ {\mathbf{v} \in \mathbb{R}^{d}:{{\| \mathbf{v} \|}} = 1} \} $, $\nabla f(\bm{\theta}, X) = \frac{1}{\left| {X} \right|}\sum\nolimits_{x\in X} \nabla f(\bm{\theta}, x)$, and $q(\bm{\theta}, X) \stackrel{\text{def}} = \nabla f(\bm{\theta}, X) - \nabla f(\bm{\theta}^*, X)$. We use $X_s$ to denote the trusted dataset at the server. 
Next, we first state the  assumptions in our theoretical analysis and then describe our theoretical results. 

\begin{assumption}
	\label{assumption_1}
	The expected loss $F(\bm{\theta})$ has $L$-Lipschitz continuous gradients and is $\mu$-strongly convex, i.e.,   $\forall \bm{\theta}, {\bm{\theta}}^{\prime} \in \Theta$, the following  inequalities hold:
	\begin{gather*}
	\left\| {\nabla F(\bm{\theta}) - \nabla F({\bm{\theta}}^{\prime}) } \right\| \le L\left\| {\bm{\theta} - {\bm{\theta}}^{\prime} } \right\|, \\
	F(\bm{\theta}) + \left\langle {\nabla F(\bm{\theta}),{\bm{\theta}}^{\prime} - \bm{\theta}} \right\rangle + \frac{\mu}{2}{\left\| {\bm{\theta}}^{\prime} - \bm{\theta} \right\|^2} \le
	F({\bm{\theta}}^{\prime}).
	\end{gather*}
\end{assumption}

\begin{assumption}
	\label{assumption_2}
	There exist constants $\alpha_1 >0$ and $\rho_1 >0$ such that for any $\mathbf{v} \in \bm{V}$,  $\left\langle {\nabla f(\bm{\theta}^*, X),\mathbf{v}} \right\rangle$ is sub-exponential. 
	That is, $\forall \left| \varphi  \right| \le 1/\rho_1$, we have: 
	\begin{equation*}
	\mathop {\sup }\limits_{\mathbf{v} \in \bm{V}} \mathbb{E}\left[ {\exp \left(    \varphi \left\langle {\nabla f(\bm{\theta}^*, X), \mathbf{v}} \right\rangle \right)  } \right]
	\le 
	e^{ \alpha_1 ^2 \varphi ^2 /2}.
	\end{equation*}
\end{assumption}

\begin{assumption}
	\label{assumption_3}
	There exist constants $\alpha_2 >0$, $\rho_2 >0$ such that for any $\mathbf{v} \in \bm{V}$, $\bm{\theta} \in \Theta$ and $\bm{\theta} \ne \bm{\theta}^*$. 
	$\left\langle {q(\bm{\theta}, X) - \mathbb{E}\left[ {q(\bm{\theta}, X)} \right],\mathbf{v}} \right\rangle / \left\| {\bm{\theta} - \bm{\theta}^*} \right\|$  is sub-exponential. 
	That is, $\forall \left| \varphi  \right| \le 1/\rho_2$, we have:
	\begin{equation*}
	\mathop {\sup }\limits_{\bm{\theta} \in \Theta ,\mathbf{v} \in \mathbf{V}} \!\!\!\!  \mathbb{E} \left[ {\exp \left( { {\varphi \left\langle {q(\bm{\theta}, X) \!-\! \mathbb{E} \left[ {q(\bm{\theta}, X)} \right],\mathbf{v}} \right\rangle }} /
		{\left\| {\bm{\theta} - \bm{\theta}^*} \right\|} \right)} \right]
	\le 
	e ^{\alpha_2 ^2 \varphi ^2 /2}.
	\end{equation*}

\end{assumption}

\begin{assumption}
	\label{assumption_4}
	For any $\beta  \in (0,1)$, there exists a constant $H>0$ such that the following inequality holds:
	\begin{align}
	 &\mathbb{P} \left\{ {\mathop {\sup }\limits_{\bm{\theta}, \bm{\theta}^{\prime} \in \Theta :\bm{\theta} \ne \bm{\theta}^{\prime}} {\left\| \frac{1}{\left| {X_s} \right|} \!\! \sum\limits_{x \in {X_s}} \! \left( \nabla f(\bm{\theta},x)  \!-\!  \nabla f(\bm{\theta}^{\prime},x) \right) \right\|} 
	\!	\le \! H {\left\| {\bm{\theta} \!-\! \bm{\theta}^{\prime}} \right\|} } \right\} \nonumber  \\
	& \qquad \ge 1 - {\beta }/{3}. \nonumber
	\end{align}	
\end{assumption}

\begin{assumption}
	\label{assumption_5}
	Clients' local training data are independent and identically distributed (i.i.d.).
	The trusted dataset held by the server and the overall training data are drawn from the same distribution, and the server delay $\tau_s=0$.
\end{assumption}

\begin{remark} {\em
	Assumption~\ref{assumption_1} is satisfied in many learning models (e.g., linear regression and quadratically regularized models). 
	Note that we only assume that the expected loss  is strongly-convex, while the empirical loss could still be non-convex. 
	Assumptions~\ref{assumption_2}-\ref{assumption_3} characterize sub-exponential properties on the gradient vectors.
	Assumption~\ref{assumption_2} is a standard assumption in the literature on convergence analysis, while
	Assumptions~\ref{assumption_2} and \ref{assumption_3} are also widely used in Byzantine-robust FL community (see, e.g., \cite{ChenPOMACS17,su2019securing,yin2018byzantine}).
	Assumption~\ref{assumption_4} is satisfied if the model/loss function is Lipschitz-smooth (e.g., regressions, neural networks).
	Assumption~\ref{assumption_5} is a sufficient condition only needed in our theoretical analysis, which characterizes the statistical relations between the server's trusted dataset and the overall training data.
Note that we only need these assumptions to provide theoretical analysis of our proposed AFLGuard, and these assumptions are commonly used in the machine learning and security communities in order to establish the convergence of the FL methods~\cite{ChenPOMACS17,yin2018byzantine,cao2020fltrust}.
In practice, some of these assumptions may not hold, e.g., clients' local training data could be non-i.i.d., trusted data held by the server and the overall training data may come from different distributions.
In Section~\ref{sec:exp}, we will first use a synthetic dataset that satisfies all assumptions to evaluate the performance of our AFLGuard. Then, we will show that AFLGuard can still effectively defend against poisoning attacks in real-world datasets and complex models when some assumptions are violated.
As a concrete example, the following lemma shows that linear regression models satisfy Assumptions~\ref{assumption_1}-\ref{assumption_4} with appropriate parameters, and the proof is shown in Appendix~\ref{sec:appendix_3}.
	}
\end{remark}	

\begin{lemma}
	\label{lemma_assum}
	Let $x_i = (\bm{u}_i, y_i)$ be the input data and define the loss function as $f(\bm{\theta}, x_i) = \frac{ \left( \left\langle \bm{u}_i, \bm{\theta}  \right\rangle  - y_i \right)^2}{2}$.
	Suppose that $y_i$ is generated by a linear regression model
	$
	y_i = \left\langle \bm{u}_i, \bm{\theta}^* \right\rangle + e_i,
	$
	where $\bm{\theta}^*$ is the unknown true model, $\bm{u}_i \sim N(0, \bm{I})$, $e_i \sim N(0,1)$ and $e_i$ is independent of $\bm{u}_i$. 
	The linear regression model satisfies Assumptions~\ref{assumption_1}-\ref{assumption_4} with the following parameters: 		
	i) Assumption~\ref{assumption_1} is satisfied with $L=1, \mu=1$; ii) Assumption~\ref{assumption_2} is satisfied with $\alpha_1 = \sqrt{2}, \rho_1 = \sqrt{2}$; iii) Assumption~\ref{assumption_3} is satisfied with $\alpha_2 = \sqrt{8}, \rho_2 = 8$; and iv) Assumption~\ref{assumption_4} is satisfied with $H = 2\text{log}(4/\beta) + 2\sqrt{d \text{log} (4/\beta)} + d$.
\end{lemma}

The following theorem shows the security of AFLGuard:

\begin{thm}
	\label{theorem_2}
	Suppose Assumptions~\ref{assumption_1}-\ref{assumption_5} are satisfied. If the global learning rate in Algorithm~\ref{AFLGuard_lip_alg} satisfies $\eta \le \frac{2}{\mu+ L}$ and each client uses one mini-batch to calculate the model update,   
	then for any number of malicious clients,
	with probability at least $1 - \beta$, 
	we have:
	\begin{align}
	\label{theorem2equ}
	\left\| \bm{\theta}^t - \bm{\theta}^* \right\| 
	\le
	\left( 1- q \right)^t \left\| \bm{\theta}^0 - \bm{\theta}^* \right\| +  4\eta \Gamma  ( \lambda +1) / q, 
	\end{align}
	where $q = 1 -  ( \sqrt{1 - \frac{2\eta \mu L }{\mu +L} }  + \eta L\lambda + 8\eta\Lambda ( \lambda +1)) $, 
		${\Gamma} = \alpha_1 \sqrt {2K_1/{\left| X_s \right|}}$,
	$\Lambda = {\alpha_2}  
	\sqrt {2(K_2 + K_3)/{\left| X_s \right|}}$, 
	$K_1 = d\log 6 + \log(3/\beta)$,
	$K_2 = d\log ({18 R}/{\alpha_2})$, $K_3 = \frac{1}{2}d\log ({\left| {X_s} \right|}/d) + \log \left( \frac{{6\alpha_2^2  \epsilon \sqrt {\left| X_s \right|} }}{\rho_2{\alpha_1}\beta } \right)$,
	$R = \max \left\{ {L,H} \right\}$,  $\epsilon>0$ is a constant, $d$ is the dimension of $\bm{\theta}$, and $\left| {X_s} \right|$ is the trusted dataset size.
\end{thm}	
\begin{proof}
	Please see Appendix~\ref{sec:appendix_2}.
\end{proof}

\begin{remark} {\em
	Our Theorem~\ref{theorem_2} shows that the convergence of AFLGuard does not require the trusted dataset size to depend on the number of model parameters, the client dataset sizes, and the number of malicious clients. 
	The trusted dataset size affects the convergence neighborhood size (the second term on the right-hand-side of Eq.~(\ref{theorem2equ})). 
	The larger the trusted dataset size, the smaller the convergence neighborhood.
	}
\end{remark}




\section{Empirical Evaluation}  \label{sec:exp}

\begin{table*}[htb!]
	\centering
	\caption{MSE and MEE of different defenses under different attacks on synthetic dataset.
		The results are in the form of “MSE / MEE”.
		“$> 1000$” means the value is larger than 1000.}
	\vspace{1mm}
	\centering
	\label{tab:Asynchronous_error_syn}
		\small 
	{
		\begin{tabular}{|c|c|c|c|c|c|c|c|}
			\hline
			& \multicolumn{1}{c|}{AsyncSGD} & \multicolumn{1}{c|}{Kardam}  & \multicolumn{1}{c|}{BASGD} & \multicolumn{1}{c|}{Zeno++} & \multicolumn{1}{c|}{AFLGuard}\\
			\hline
			No attack & 0.03  / 0.18   &  0.03 / 0.18  & 0.09  / 1.43  & 0.03  / 0.40 &   0.03  / 0.18  \\
			\hline
			LF attack &  21.05   / 25.75   &  0.04 / 0.60  & 16.71 / 22.70    & 0.03  / 0.40 &    0.03  / 0.18  \\
			\hline
			Gauss attack &  0.78 / 4.82    &  0.03 / 0.36  & 0.85 / 5.32  &  0.03  / 0.40  &  0.03  / 0.18  \\
			\hline
			GD attack &  “$> 1000$” / “$> 1000$”     &   30.14 / 30.65  & “$> 1000$” / “$> 1000$” & 0.03  / 0.40 &   0.03  / 0.18   \\
			\hline
			Adapt attack & “$> 1000$” / “$> 1000$”    &  “$> 1000$” / “$> 1000$”    & “$> 1000$” / “$> 1000$”   & 0.03  / 0.42  &  0.03  / 0.18   \\
		\hline
		\end{tabular}%
	}
\end{table*}%

\begin{table}[htb!]
	\centering
	\caption{Test error rates and attack success rates of different defenses under different attacks on real-world datasets. The results of BD attack are in the form of “test error rate / attack success rate”.}
	\centering
	\addtolength{\tabcolsep}{-2.8pt}
	\label{tab:Asynchronous_error}
		\small 
\begin{tabular}{c}
\textbf{\footnotesize (a) MNIST}\\
\end{tabular}\\
	{
		\begin{tabular}{|c|c|c|c|c|c|c|c|}
			\hline
			& \multicolumn{1}{c|}{AsyncSGD} & \multicolumn{1}{c|}{Kardam}  & \multicolumn{1}{c|}{BASGD} & \multicolumn{1}{c|}{Zeno++} & \multicolumn{1}{c|}{AFLGuard}\\
			\hline
			No attack &  0.05     &  0.12 & 0.19   & 0.08  &   0.06   \\
			\hline
			LF attack &  0.09     &  0.15  & 0.26    & 0.09 &   0.07  \\
			\hline
			Gauss attack &  0.91    &  0.39  & 0.27  &   0.09    & 0.07 \\
			\hline
			GD attack &  0.90     &   0.90  & 0.89  & 0.09 &   0.07  \\
			\hline
			BD attack &  0.90 / 1.00     &    0.91 / 1.00   &  0.91 / 1.00    &  0.09 / 0.01  &   0.07 / 0.01   \\
			\hline
			Adapt attack & 0.91    &   0.91  &  0.90  & 0.10  & 0.07  \\
			\hline
		\end{tabular}%
	}
	\\
		\vspace{0.05in}
\begin{tabular}{c}
\textbf{\footnotesize (b) Fashion-MNIST}\\
\end{tabular}\\
	{
		\begin{tabular}{|c|c|c|c|c|c|c|c|}
			\hline
			& \multicolumn{1}{c|}{AsyncSGD} & \multicolumn{1}{c|}{Kardam}  & \multicolumn{1}{c|}{BASGD} & \multicolumn{1}{c|}{Zeno++} & \multicolumn{1}{c|}{AFLGuard}\\
			\hline
			No attack &  0.15     &   0.29  & 0.24  & 0.26  &   0.17    \\
			\hline
			LF attack &  0.19     &  0.29 & 0.24   & 0.29 &   0.21  \\
			\hline
			Gauss attack &   0.90    &   0.29 & 0.35  &   0.28    & 0.19 \\
			\hline
			GD attack &  0.90     &   0.90  & 0.90  & 0.29 &   0.21  \\
			\hline
			BD attack &  0.90 / 1.00     &   0.90 / 1.00  &  0.90 / 1.00  &  0.29 / 0.05  &   0.20 / 0.04   \\
			\hline
			Adapt attack & 0.90    &   0.90  &   0.90  & 0.29  & 0.21  \\
		\hline
		\end{tabular}%
	}
	\\
	\vspace{0.05in}
\begin{tabular}{c}
\textbf{\footnotesize (c) HAR} \\
\end{tabular}\\
	{
	\begin{tabular}{|c|c|c|c|c|c|c|c|}
		\hline
		& \multicolumn{1}{c|}{AsyncSGD} & \multicolumn{1}{c|}{Kardam}  & \multicolumn{1}{c|}{BASGD} & \multicolumn{1}{c|}{Zeno++} & \multicolumn{1}{c|}{AFLGuard}\\
		\hline
		No attack & 0.05     &  0.06  & 0.07  & 0.06  &  0.05    \\
		\hline
		LF attack & 0.19     &  0.22  &0.08  & 0.08  &  0.05    \\
		\hline
		Gauss attack & 0.30     &  0.23  &0.24  & 0.07  &  0.05    \\
		\hline
		GD attack & 0.83     &  0.48   &0.67  &  0.08  &  0.05    \\
		\hline
		BD attack &  0.18 / 0.47     &    0.17 / 0.02  & 0.41 / 0.28  &  0.07 / 0.01  &   0.05 / 0.01   \\
		\hline
		Adapt attack & 0.93    &   0.52  &  0.90  & 0.08  & 0.05  \\
		\hline
	\end{tabular}%
	}
	\\
	\vspace{0.05in}
\begin{tabular}{c}
\textbf{\footnotesize (d) Colorectal Histology MNIST} \\
\end{tabular}\\
	{
	\begin{tabular}{|c|c|c|c|c|c|c|c|}
			\hline
			& \multicolumn{1}{c|}{AsyncSGD} & \multicolumn{1}{c|}{Kardam}  & \multicolumn{1}{c|}{BASGD} & \multicolumn{1}{c|}{Zeno++} & \multicolumn{1}{c|}{AFLGuard}\\
			\hline
			No attack &  0.21    &  0.28   & 0.29   &  0.31 &  0.22   \\
			\hline
			LF attack &  0.29     & 0.37  & 0.40  & 0.39 &  0.23 \\
			\hline
			Gauss attack &  0.65    &  0.44   & 0.61   &   0.43    & 0.22 \\
			\hline
			GD attack & 0.87      & 0.68   & 0.87  & 0.39 &  0.32 \\
			\hline
			BD attack &  0.75 / 0.84     &    0.67 / 0.02  &  0.85 / 0.84   &  0.44 / 0.02  &   0.27 / 0.02   \\
			\hline
			Adapt attack & 0.88    &   0.88  &  0.88  & 0.64  & 0.33  \\
			\hline
		\end{tabular}%
	}
	\\
	\vspace{0.05in}
\begin{tabular}{c}
\textbf{\footnotesize (e) CIFAR-10} \\
\end{tabular}\\
	{
		\begin{tabular}{|c|c|c|c|c|c|c|c|}
			\hline
			& \multicolumn{1}{c|}{AsyncSGD} & \multicolumn{1}{c|}{Kardam}  & \multicolumn{1}{c|}{BASGD} & \multicolumn{1}{c|}{Zeno++} & \multicolumn{1}{c|}{AFLGuard}\\
			\hline
			No attack & 0.26   &   0.29  & 0.47  & 0.41  &   0.26  \\
			\hline
			LF attack &  0.40  &  0.52 & 0.54 & 0.53 & 0.34  \\
			\hline
			Gauss attack &  0.88    &   0.63  & 0.81  &  0.52     & 0.33 \\
			\hline
			GD attack & 0.90      &  0.90  & 0.90 & 0.60 &  0.30 \\
			\hline
			BD attack &  0.76 / 0.99     &    0.82 / 1.00   & 0.74 / 0.98   &  0.49 / 0.06  &   0.29 / 0.01   \\		
			\hline
			Adapt attack & 0.90    &   0.90  &  0.90  & 0.82  & 0.36  \\
			\hline
		\end{tabular}%
	}
\end{table}%

\subsection{Experimental Setup}

\subsubsection{Compared Methods}

We compare our AFLGuard  with the following asynchronous methods:

\myparatight{1)  AsyncSGD~\cite{zheng2017asynchronous}} In AsyncSGD, the server updates the global model  according to Algorithm~\ref{AsynSGD_alg} upon receiving a model update from any client.

\myparatight{2) Kardam~\cite{damaskinos2018asynchronous}} 
In Kardam, the server keeps an empirical Lipschitz coefficient  for each client, and filters out potentially malicious model updates based on the Lipschitz filter.

\myparatight{3) BASGD~\cite{yang2021basgd}} In BASGD, the server holds several buffers. Upon receiving a model update from any client, the server stores it into one of these buffers according to a mapping table. 
When all buffers are non-empty, the server  computes the average of model updates in each buffer,  takes the median of all buffers, and uses it to update the global model.

\myparatight{4) Zeno++~\cite{xie2020zeno++}} 
In Zeno++, the server has a trusted dataset. Upon receiving a client's model update, the server computes a server model update based on the trusted dataset. If the cosine similarity between the server model update and the client's model update is positive, then the server normalizes the client's model update to have the same magnitude as the server model update and uses the normalized model update to update the global model.

\subsubsection{Datasets}

We evaluate AFLGuard and the compared methods using one synthetic dataset  and five real-world datasets (MNIST, Fashion-MNIST, Human Activity Recognition (HAR), Colorectal Histology MNIST, CIFAR-10). The synthetic dataset is for linear regression, which satisfies the Assumptions~\ref{assumption_1}-\ref{assumption_4} in Section~\ref{sec:sec_ana}
and is used to validate our theoretical results. Other datasets are used to train complex models, which aim to show the effectiveness of AFLGuard even if the Assumptions~\ref{assumption_1}-\ref{assumption_4}  are not satisfied. 
The details of these datasets are shown in Appendix~\ref{sec:appendix_datasets} due to limited space.

\subsubsection{Poisoning Attacks}
We use the following poisoning attacks in our experiments.

\myparatight{1) Label flipping (LF) attack~\cite{yin2018byzantine}} 
In the LF attack, the label $y$ of each training example in the malicious clients is replaced by $C-1-y$, where $C$ is the total number of classes. 
For instance, for the MNIST dataset, digit ``1'' is replaced by digit ``8''.

\myparatight{2) Gaussian (Gauss) attack~\cite{blanchard2017machine}}
In the Gauss attack, each model update from malicious clients is drawn from a zero-mean Gaussian distribution (we set the standard deviation to 200).

\myparatight{3) Gradient derivation (GD) attack~\cite{fang2020local}}
In the GD attack adapted from~\cite{fang2020local}, a malicious client computes a model update based on its local training data and then scales it by a negative constant ($-10$ in our experiments) before sending it to the server. 

\myparatight{4) Backdoor (BD) attack~\cite{gu2017badnets,bagdasaryan2020backdoor,cao2020fltrust}}
BD attack is a targeted poisoning attack. 
We use the same strategy in~\cite{gu2017badnets} to embed the trigger in MNIST, Fashion-MNIST and Colorectal Histology MNIST datasets. 
Following~\cite{cao2020fltrust}, the target label is set to “WALKING UPSTAIRS” and the trigger is generated by setting every 20th feature to 0 for the HAR dataset.
For the CIFAR-10 dataset, the target label is set to “bird” and we use the same pattern trigger as suggested in~\cite{bagdasaryan2020backdoor}.

\myparatight{5) Adaptive (Adapt) attack} 
In~\cite{fang2020local}, a general adaptive attack framework is proposed to attack FL with any aggregation rule. 
We apply this attack framework to construct an adaptive attack to our AFLGuard. In particular, the attack framework is designed for synchronized FL, in which the server aggregates model updates from multiple clients to update the global model. The key idea is to craft model updates at the malicious clients such that the aggregated model update deviates substantially from the before-attack one. To apply this general attack framework to  AFLGuard, we assume that the server accepts or rejects a client's model update based on AFLGuard and computes the average of the accepted model updates. Then, we craft the model updates on the malicious clients based on the attack framework.

\subsubsection{Evaluation Metrics}
For the synthetic dataset, we use the following two evaluation metrics since it is a regression problem:
i) \emph{Mean Squared Error (MSE)}: MSE is computed as $\text{MSE} = \frac{1}{N_t} \sum\nolimits_{i=1}^{N_t} {(\hat{y}_i - y_i)^2} $, where $\hat{y}_i$ is the predicted value, $y_i$ is the true value, and $N_t$ is the number of testing examples;
ii) \emph{Model Estimation Error (MEE)}: MEE is computed as $\text{MEE} =\| \hat{\bm{\theta}} - \bm{\theta}^* \|_2$, where $\hat{\bm{\theta}}$ is the learnt model and $\bm{\theta}^*$ is the true model. MEE is commonly used in measuring the performance of linear regression~\cite{gupta2019byzantine,turan2021robust}.
The five real-world datasets represent classification tasks, and we consider the following two evaluation metrics: 1) \emph{test error rate}, which is the fraction of clean testing examples that are misclassified; and  2) \emph{attack success rate}, which is the fraction of trigger-embedded testing inputs that are predicted as the attacker-chosen target label. Note that attack success rate is only applicable for targeted poisoning attack (i.e., BD attack in our experiments). 
The smaller the error (MSE, MEE, or test error rate) and attack success rate, the better the defense. 
Note that we do not consider targeted poisoning attacks on synthetic dataset since there are no such attacks designed for linear regression.

\subsubsection{Parameter Setting}
We assume 100 clients ($n=100$) for synthetic, MNIST, and Fashion-MNIST datasets, and 40 clients ($n=40$) for Colorectal Histology MNIST and CIFAR-10 datasets. 
The HAR dataset is collected from smartphones of 30 real-world users, and each user is considered as a client.
Thus, there are 30 clients ($n=30$) in total for HAR. 
By default, we assume 20\% of the clients are malicious. 
We train a convolutional neural network (CNN) on MNIST and Fashion-MNIST datasets, and its architecture is shown in Table~\ref{cnn_arch} in Appendix.
We train a logistic regression classifier on HAR dataset.
We train a ResNet-20~\cite{he2016deep} model for Colorectal Histology MNIST and CIFAR-10 datasets.
We set 2,000, 2,000, 6,000, 1,000, 20,000 and 20,000 iterations for synthetic, MNIST, Fashion-MNIST, HAR, Colorectal Histology MNIST and CIFAR-10 datasets, respectively.
The batch sizes for the six datasets are 16, 32, 64, 32, 32 and 64, respectively.
The learning rates are set to 1/1600, 1/320 for synthetic and HAR datasets, respectively; and are set to 1/3200 for the other four datasets. We use different parameters for different datasets because they have different data statistics. 
In the synthetic dataset, we assume the clients' local training data are i.i.d.
However, the local training data non-i.i.d. across clients in the five real-world datasets.
In particular, we use the approach in~\cite{fang2020local} to simulate the non-i.i.d. setting. 
The non-i.i.d. degree is set to 0.5 for MNIST, Fashion-MNIST, Colorectal Histology MNIST, and CIFAR-10 datasets. 
Note that each user is a client in HAR dataset, and thus the clients' local training data are already heterogeneous for HAR.

\begin{figure*}[!t]
	\centering
	\includegraphics[scale = 0.5]{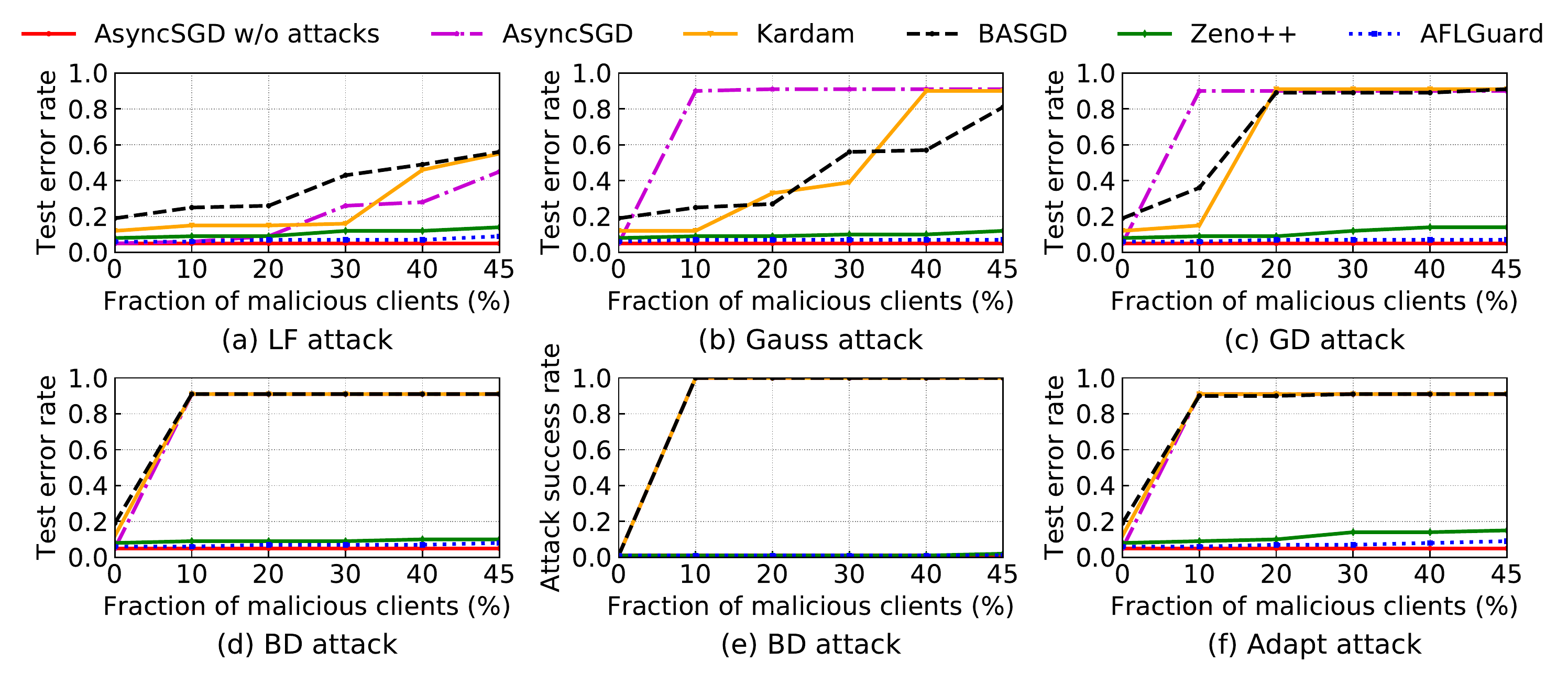}	
	\caption{Test error rates and attack success rates of different defenses under different attacks with different fraction of malicious clients on MNIST dataset.}
	\label{attack_size_single}
	\vspace*{-13pt}
\end{figure*}

In AFLGuard, the trusted dataset size is set to 100 for all six datasets. 
By default, for the synthetic data, we assume that the trusted dataset held by the server and the overall training data are generated from the same distribution. 
For the real-world datasets, we do not make this assumption. 
We will empirically show that our method works well even if the distribution of trusted data deviates  from that of the overall training data, i.e., there exists a \textit{distribution shift (DS)} between these two datasets.
The larger the DS, the larger the deviation between the trusted and overall training datasets.
In our experiments, we simulate DS in the following way:
a fraction of samples in the trusted dataset are drawn from one particular class (the first class in our experiments) of the overall training data, and the remaining samples in the trusted dataset are drawn from the remaining classes of the overall training data uniformly at random.
We use this fraction value as a proxy for DS. 
Note that when the trusted and overall training datasets are drawn from the same distribution, DS is equal to $1/C$, where $C$ is the total number of classes. 
By default, we set DS to 0.5 for the five real-world datasets.

\looseness-1  In our experiments, we use a separate validation dataset to tune the parameter $\lambda$  in  AFLGuard. 
Note that this validation dataset is different from the trusted dataset held by the server.
We use the validation dataset to tune the hyperparameter of AFLGuard, while the server in AFLGuard uses the trusted dataset to filter out potential malicious information. The size of the validation dataset is 200. The validation data and the overall training data (the union of the local training data of all clients) are from the same distribution. For example, there are 10 classes in MNIST dataset. We  sample 20 training examples from each class of the overall training data uniformly at random. After fine tuning the parameter, we set $\lambda=1.5$ for synthetic, MNIST, and HAR datasets, and $\lambda=1.8$ for the other three datasets.
The server updates $\bm{g}_s^{t-\tau_s}$ every 10 ($\tau_s=10$) iterations by default. 
We use the approach in~\cite{xie2020zeno++} to simulate asynchrony. 
We sample client delay from the interval $[0, \tau_{\text{max}}]$ uniformly at random, where $\tau_{\text{max}}$
is the maximum client delay. 
We set $\tau_{\text{max}}=10$ by default.

\begin{figure*}[!t]
	\centering
	\includegraphics[scale = 0.49]{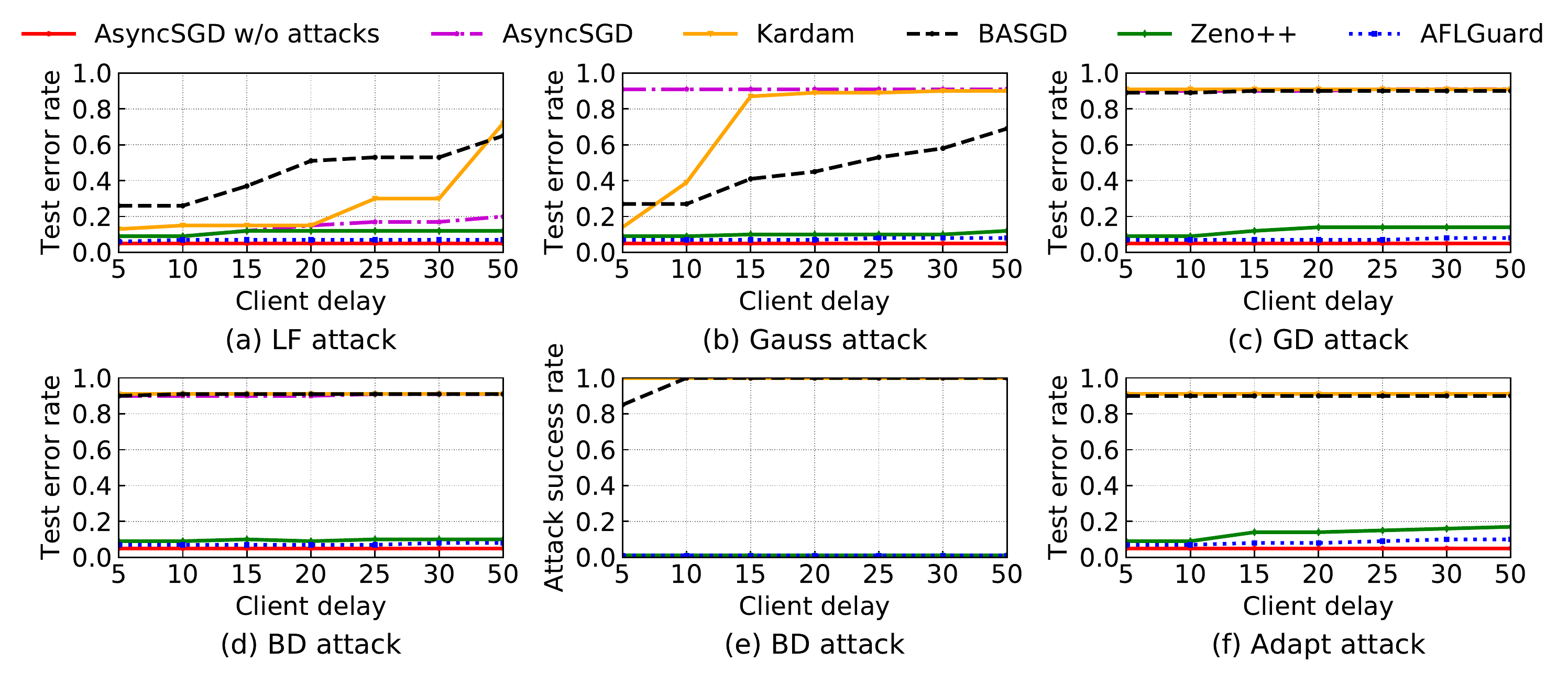}
	\caption{Test error rates and attack success rates of different defenses under different attacks with different client delays on MNIST dataset.}
	\label{client_delay_single}
\vspace*{-10pt}\end{figure*}

\begin{figure*}[!t]
	\centering
	\includegraphics[scale = 0.49]{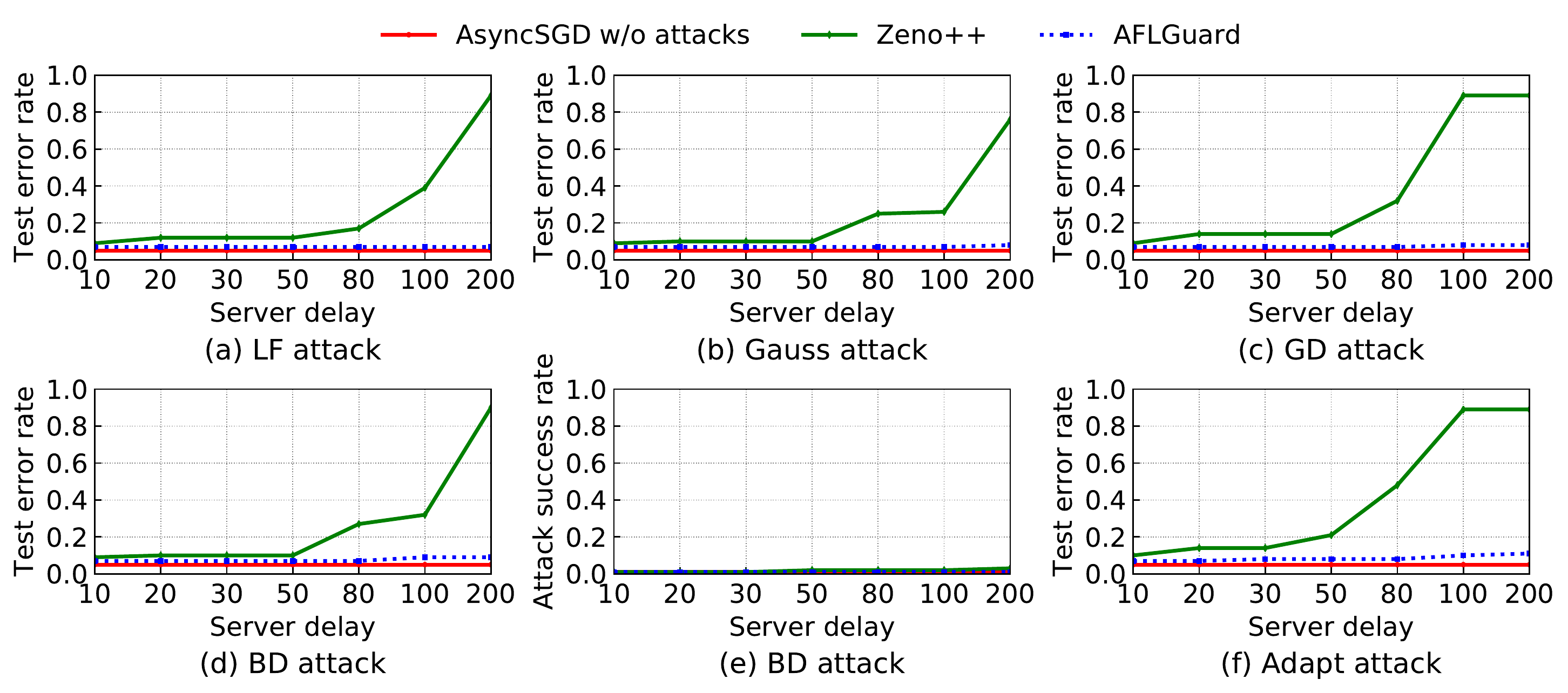}
	\caption{Test error rates and attack success rates of Zeno++ and AFLGuard under different attacks with different server delays on MNIST dataset.}
	\label{server_delay_single}
\vspace*{-12pt}\end{figure*}

\subsection{Experimental Results}

\begin{figure}[!t]
	\centering
	\includegraphics[scale = 0.230]{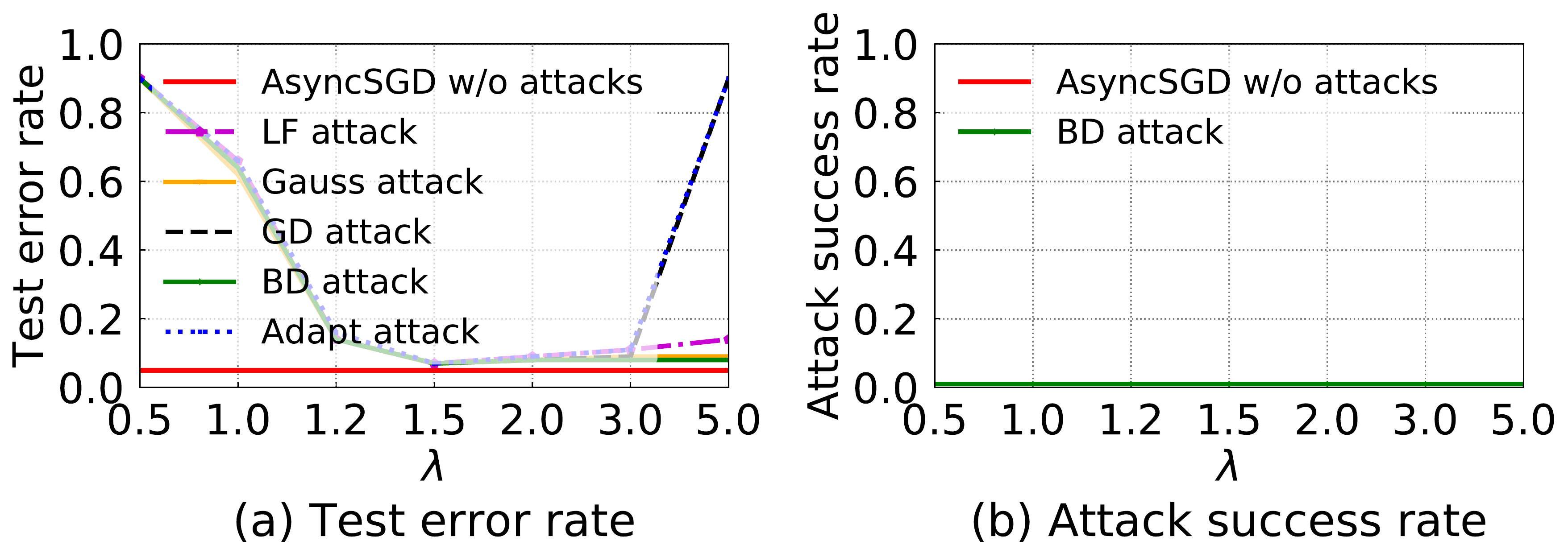}
	\caption{Test error rates and attack success rates of AFLGuard under different attacks with different $\lambda$ on MNIST dataset.}
	\label{alfguard_lambda}
	\vspace*{-12pt}
\end{figure}

\myparatight{AFLGuard is Effective} 
We first show results on the linear regression model for synthetic dataset, which satisfies Assumptions~\ref{assumption_1}-\ref{assumption_4} in Section~\ref{sec:sec_ana}
to support our theoretical results.
The MSE and  MEE of different methods under different attacks on synthetic dataset are shown in Table~\ref{tab:Asynchronous_error_syn}. 
We observe that AFLGuard is robust in both non-adversarial and adversarial settings. In particular,  the MSE and MEE of AFLGuard under various attacks are the same as those of AsyncSGD without attacks. 
Moreover, we also observe that AFLGuard outperforms the compared methods.
For instance, the MSE and MEE of BASGD are both larger than 1,000 under the GD and Adapt attacks.

Next, we show results on the five real-world datasets. 
The test error rates and attack success rates of different methods under different attacks are shown in Table~\ref{tab:Asynchronous_error}. 
``No attack” in Table~\ref{tab:Asynchronous_error} represents the test error rate without any attacks.
For the untargeted poisoning attacks (LF attack, Gauss attack, GD attack, and Adapt attack), the results are the test error rates;
and for the targeted poisoning attacks (BD attack), the results are in the form of “test error rate / attack success rate”.
We note that only using the trusted data held by the server to update the global model can not achieve satisfactory accuracy.
For instance, the test error rate is 0.21 when we only use the trusted data of the server to update the global model on the MNIST dataset.
We also remark that our asynchronous AFLGuard algorithm achieves a performance similar to its synchronous counterpart. For instance,
on MNIST, the test error rates of synchronous AFLGuard under LF, Gauss, and GD attacks are all 0.05.

\begin{figure}[!t]
	\centering
	\includegraphics[scale = 0.230]{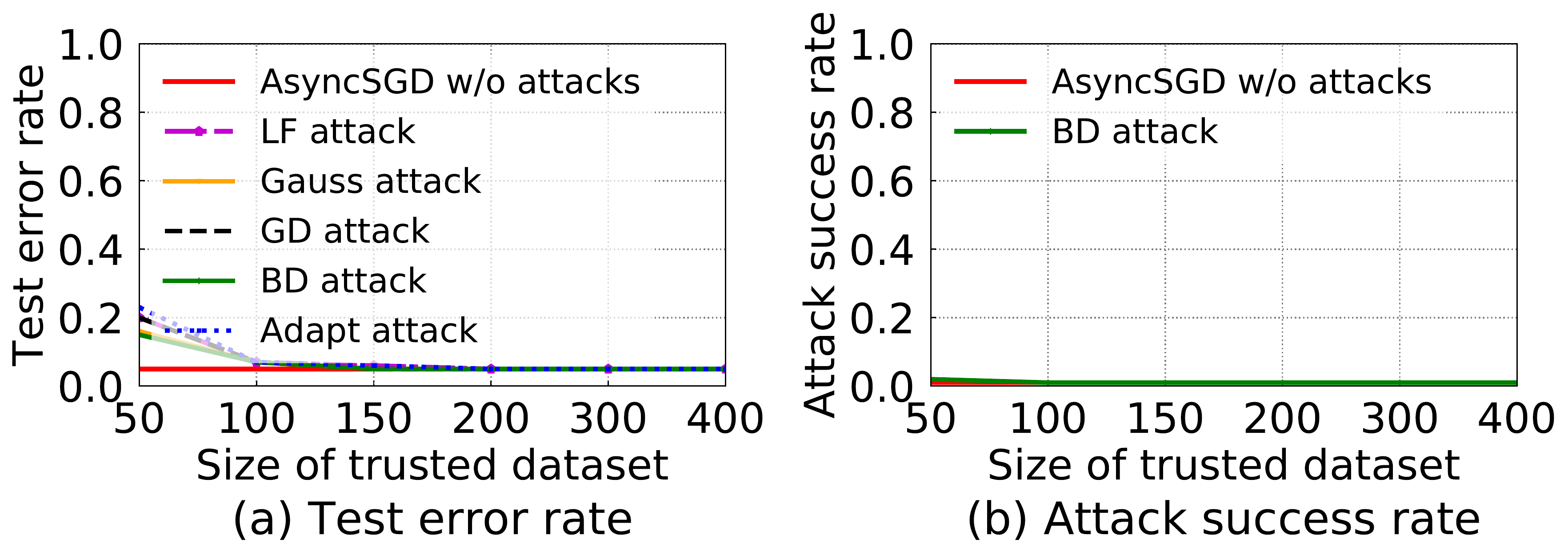}
	\caption{Test error rates and attack success rates of AFLGuard under different attacks with different size of trusted dataset on MNIST dataset.}
	\label{alfguard_trust_size}
\vspace*{-10pt}
\end{figure}

First, we observe that AFLGuard is effective in non-adversarial settings. 
When there are no malicious clients,  AFLGuard  has similar test error rate as AsyncSGD.
For instance, on MNIST, the test error rates without attacks are respectively 0.05 and 0.06 for AsyncSGD and AFLGuard, while the test error rates are respectively 0.12 and 0.19 for  Kardam and BASGD. 
Second, AFLGuard is robust against various poisoning attacks and outperforms all baselines.
For instance, the test error rate of Kardam increases to 0.90 under the GD attack on the MNIST and Fashion-MNIST datasets, while the test error rates are 0.07 and 0.21 for AFLGuard under the same setting. Likewise, the attack success rates of AFLGuard are at most 0.04 for all real-world datasets, while   the attack success rates of AsyncSGD, Kardam, and BASGD are high. 
Note that in Table~\ref{tab:Asynchronous_error_syn}, we use synthetic data that satisfies Lemma~\ref{lemma_assum} to evaluate the performance of our AFLGuard.
Since the variance of the synthetic data is small, Zeno++ and AFLGuard have similar MSE and MEE.  However, Table 2 shows that, for real-world datasets, AFLGuard significantly outperforms Zeno++.

\myparatight{Impact of the Fraction of Malicious Clients}
Fig.~\ref{attack_size_single} illustrates the test error rates and attack success rates of different methods under different attacks on the MNIST dataset, when the fraction of
malicious clients increases from 0 to 45\%.
Note that Fig.~\ref{attack_size_single}(e) shows the attack success rates of different defenses under BD attack, while other figures are the test error rates of different defenses under untargeted and targeted poisoning attacks.
We observe that AFLGuard achieves a test error rate similar to that of AsyncSGD without attacks, when 45\% of clients are malicious.
This shows that AFLGuard is robust against a large fraction of malicious clients.

\begin{table*}[htbp]
	\centering
	\caption{Test error rates and attack success rates of Zeno++ and AFLGuard under different attacks with different distribution shifts (DSs) between the trusted data and overall training data on MNIST dataset. The results of BD attack are in the form of “test error rate / attack success rate”.}
	\vspace{1mm}
	\label{bias_single}%
		\small 
	{
		\begin{tabular}{|c|c|c|c|c|c|c|c|c|c|c|}
			\hline
			DS   & \multicolumn{2}{c|}{0.1} & \multicolumn{2}{c|}{0.5} & \multicolumn{2}{c|}{0.6} & \multicolumn{2}{c|}{0.8} & \multicolumn{2}{c|}{1.0} \\
			\hline
			Attack 
			&\tabincell{c}{Zeno++}   & \tabincell{c}{AFLGuard}  
			& \tabincell{c}{Zeno++}   & \tabincell{c}{AFLGuard}
			& \tabincell{c}{Zeno++}   & \tabincell{c}{AFLGuard}   &\tabincell{c}{Zeno++}   & \tabincell{c}{AFLGuard}   & \tabincell{c}{Zeno++}   & \tabincell{c}{AFLGuard} \\
			\hline
			No attack & 0.05 & 0.05  & 0.08  &  0.06 & 0.10  & 0.06  & 0.55  & 0.06  & 0.88    &  0.11 \\
			\hline
			LF attack & 0.07 & 0.05 &   0.09    & 0.07       & 0.12      &  0.07     &    0.86   &  0.07    &  0.89     &  0.11 \\
			\hline
			Gauss attack   & 0.07 & 0.05  &   0.09    &  0.07     &  0.12     &    0.07   &  0.59     &   0.07    &   0.89    & 0.12 \\
			\hline
			GD attack  & 0.07 & 0.06  &   0.09    &  0.07      & 0.12    &  0.07     & 0.78      &  0.08     &   0.89    &  0.12 \\
			\hline
			BD attack  & 0.06 / 0.01 & 0.05 / 0.01  & 0.09 / 0.01   & 0.07 / 0.01   & 0.11 / 0.01  & 0.07 / 0.01 & 0.55 / 0.03 & 0.08 / 0.01   & 0.90 / 0.01 & 0.11 / 0.01 \\
			\hline
			Adapt attack  & 0.07 & 0.06  &  0.10    & 0.07      & 0.12   & 0.08    & 0.88    & 0.10   &   0.90    & 0.12 \\
			\hline
		\end{tabular}%
	}
\vspace*{-10pt}\end{table*}%

\myparatight{Impact of the Number of Clients}
Fig.~\ref{attack_client_number} in Appendix shows the results of different defenses under different attacks, when the total number of clients $n$ varies from 50 to 500. The fraction of malicious clients is set to 20\%. 
We observe that our  AFLGuard can  effectively defend against various poisoning attacks for different number of clients.
In particular, AFLGuard under attacks achieves test error rates similar to AsyncSGD without attack.

\myparatight{Impact of the Client Delay}
A challenge and key feature in asynchronous FL is the delayed client model updates.
In this experiment, we investigate the impact of the maximum client delays $\tau_{\text{max}}$ on the test error rate of different defenses under different attacks on the MNIST dataset, where the server delay is set to 10. 
The results are shown in Fig.~\ref{client_delay_single}.
We observe that AFLGuard is insensitive to the delays on the client side, and the test error rates remain almost unchanged when the client delay varies from 5 to 50. 
However,  Kardam and BASGD  are highly sensitive to client delays. 
For example, under the Gauss attack, the test error rate of Kardam increases from 0.14 to 0.39 when the client delay increases from 5 to 10. 
Moreover, under the GD and Adapt attacks, the test error rates of  Kardam and BASGD  are both 0.90 when the client delay is only 5.

\myparatight{Impact of the Server Delay}
In both Zeno++ and our  AFLGuard, the server uses server model update.
In this experiment, we investigate the impact of server delays on the performance of Zeno++ and AFLGuard  under different attacks, where the maximum client delay is set to 10. 
The results are shown in Fig.~\ref{server_delay_single}.  We observe that AFLGuard can effectively defend against various poisoning attacks with different server delays. 
 AFLGuard under attacks  has test error rates similar to those of AsyncSGD under no attacks when the server delay ranges from 10 to 200. 
However,  Zeno++  is highly sensitive to server delays. 
For instance,  Zeno++  can only resist the Adapt attack up to 80 server delays.

\myparatight{Impact of $\lambda$}
Fig.~\ref{alfguard_lambda} shows the test error rates and attack success rates of AFLGuard under different attacks with different $\lambda$ values on the MNIST dataset.
We observe that if  $\lambda$ is too small (e.g., $\lambda = 0.5$), the test error rate of AFLGuard is large since the server rejects many benign model updates. 
When  $\lambda$ is large (e.g., $\lambda = 5.0$), the test error rates of AFLGuard under the GD and Adapt attacks are large. This is because the server falsely accepts some  model updates from malicious clients.

\myparatight{Impact of the Trusted Dataset}
The trusted dataset can be characterized by its size and distribution. Therefore, we explore the impact of both its size and distribution. 
Fig.~\ref{alfguard_trust_size} shows the results of AFLGuard under different attacks on the MNIST dataset, when the size of the trusted dataset  increases from
50 to 400 (other parameters are set to their default values). 
We find that AFLGuard only requires a small trusted dataset (e.g., 100 examples) to defend against different attacks.

Table~\ref{bias_single} shows the results of Zeno++ and AFLGuard  under different attacks when the DS value between the trusted data and overall training data varies on the MNIST dataset.
The results on the other four real-world datasets are shown in Table~\ref{bias_single_app} in Appendix. 
Note that, for the synthetic data, we assume that the trusted data  and the overall training data are generated from the same distribution. 
Thus, there is no need to study the impact of DS on synthetic data. First, we observe that AFLGuard outperforms Zeno++ across different DS values in most cases, especially when DS is large. This is because Zeno++  classifies a client's model update as benign if it is not negatively correlated with the (delayed) server model update. 
However, when the trusted dataset deviates substantially from the overall training dataset, it is very likely that the server model update  and the  model updates from benign clients are not positively correlated. 
Second, AFLGuard outperforms Zeno++ even if the trusted data has the same distribution as that of overall training data (corresponding to DS being 0.1 for MNIST, Fashion-MNIST and CIFAR-10 datasets, 0.167 for HAR dataset, and 0.125 for Colorectal Histology MNIST dataset). 
Third, AFLGuard can tolerate a large DS value, which means that AFLGuard does not require the trusted dataset to have similar distribution with the overall training data.




\section{Conclusion and Future Work}
\label{sec:conclusion}

In this paper, we propose a   Byzantine-robust asynchronous FL  framework called AFLGuard to defend against poisoning attacks in asynchronous FL. 
In AFLGuard, the server holds a small and clean trusted dataset to assist the filtering of model updates from malicious clients.
We theoretically analyze the security guarantees  of AFLGuard. 
We extensively evaluate  AFLGuard  against state-of-the-art and adaptive poisoning attacks on one synthetic and five real-world datasets. 
Our results show that  AFLGuard  effectively mitigates  poisoning attacks and outperforms existing Byzantine-robust asynchronous FL methods. 
One interesting future work is to investigate the cases where the server has no knowledge of the training data domain.

\begin{acks}
	We thank the anonymous reviewers and shepherd Briland Hitaj for their constructive comments. This work was supported by NSF grant CAREER CNS-2110259, CNS-2112471, CNS-2102233, CNS-2131859, CNS-2112562, CNS-2125977 and CCF-2110252, as well as ARO grant No. W911NF2110182. 
\end{acks}


\balance
\bibliographystyle{ACM-Reference-Format}
\bibliography{refs}


\appendix


\section{Appendix}

\subsection{Proof of Lemma~\ref{lemma_assum}} \label{sec:appendix_3}

The proof of Lemma~\ref{lemma_assum} is mainly from~\cite{ChenPOMACS17, su2019securing}.
We first check Assumption~\ref{assumption_1}.
Consider the linear regression model defined in Lemma~\ref{lemma_assum}, that is 
$
y_i = \left\langle \bm{u}_i, \bm{\theta}^* \right\rangle + e_i,
$
where $\bm{\theta}^*$ is the unknown true model parameter, $\bm{u}_i \sim N(0, \bm{I})$, $e_i \sim N(0,1)$, $e_i$ is independent of $\bm{u}_i$. 
The population risk of~(\ref{eqn_ERM}) is given by
$
\min_{\bm{\theta} } \frac{1}{2} \left\|  \bm{\theta} - \bm{\theta}^* \right\|^2 + \frac{1}{2}.
$
$ F(\bm{\theta}) \triangleq \mathbb{E} \left[ f(\bm{\theta}, X) \right] = \mathbb{E}\left[ \frac{1}{2} (\left\langle \bm{u}, \bm{\theta} \right\rangle -y )^2    \right] =\mathbb{E} \left[ \frac{1}{2} (\left\langle \bm{u}, \bm{\theta} \right\rangle - \left\langle \bm{u}, \bm{\theta}^* \right\rangle -e )^2   \right] = \frac{1}{2} \left\|  \bm{\theta} - \bm{\theta}^* \right\|^2 + \frac{1}{2} $.
Then the gradient of population risk is $\nabla F(\bm{\theta}) = \bm{\theta} - \bm{\theta}^*$.
We can see that the population risk $F(\cdot)$ is  $L$-Lipschitz continuous with $L=1$, and $\mu$-strongly convex with $\mu=1$.

We then check Assumption~\ref{assumption_2}.
Let $\bm{u} \sim N(0, \bm{I})$, $e \sim N(0,1)$ and $e$ is independent of $\bm{u}$, then one has $\nabla f(\bm{\theta}, X) = \bm{u} \left\langle \bm{u}, \bm{\theta}-\bm{\theta}^* \right\rangle -\bm{u} e$.
Let $\mathbf{v} \in \bm{V}$ be the unit vector, we further have that:
\begin{align}
\label{grad_times_v}
\left\langle \nabla f(\bm{\theta}^*, X), \mathbf{v} \right\rangle = -e \left\langle \bm{u}, \mathbf{v} \right\rangle.
\end{align}

Since $\bm{u} \sim N(0, \bm{I})$, $\mathbf{v}$ is the unit vector and $\bm{u}$ is independent of $e$, so we have $\left\langle \bm{u}, \mathbf{v} \right\rangle \sim N(0,1)$ and $\left\langle \bm{u}, \mathbf{v} \right\rangle$ is independent of $e$.
According to the standard conditioning argument, for $\varphi^2 \le 1$, one has:
\begin{align}
& \mathbb{E}\left[ \text{exp} \left( \varphi  \left\langle  \nabla f(\bm{\theta}^*, X),  \mathbf{v} \right\rangle   \right)  \right] 
\stackrel{(a)} = \mathbb{E}\left[ \text{exp}  \left( - \varphi e  \left\langle \bm{u}, \mathbf{v} \right\rangle  \right)  \right]  \nonumber \\
&= \mathbb{E}\left[ \mathbb{E}\left[   \text{exp}  \left( - \varphi y \left\langle \bm{u}, \mathbf{v} \right\rangle  \right)  | \varphi=y  \right]    \right]  \nonumber \\
& \stackrel{(b)}= \mathbb{E}\left[  \text{exp}  \left(  \varphi^2 e^2/2 \right)  \right] 
\stackrel{(c)}=  \left(  1-\varphi^2  \right) ^{-1/2} 
\stackrel{(d)} \le e^{\varphi^2},
\end{align}
where $(a)$ is because of Eq.~(\ref{grad_times_v}); $(b)$ is obtained by applying the moment generating function of Gaussian distribution; $(c)$ is because for the moment generating function of $\chi^2$ distribution, we have $\mathbb{E}\left[ \text{exp}  \left( t \varphi^2 \right) \right] =\left(  1-2t  \right) ^{-1/2} $ for $t<1/2$; $(d)$ is due to the fact that $1-\varphi^2 \ge e^{-2\varphi^2}$ for $ |\varphi| \le 1/\sqrt{2}$.
Therefore, Assumption~\ref{assumption_2} holds when $\alpha_1 = \sqrt{2}$ and $\rho_1 = \sqrt{2}$.

Next, we check Assumption~\ref{assumption_3}.
As $\nabla f(\bm{\theta}, X) = \bm{u} \left\langle \bm{u}, \bm{\theta}-\bm{\theta}^* \right\rangle -\bm{u} e$, $\nabla f(\bm{\theta}^*, X)=-\bm{u} e$,
so $q(\bm{\theta}, X) = \nabla f(\bm{\theta}, X) - \nabla f(\bm{\theta}^*, X)= \bm{u} \left\langle \bm{u}, \bm{\theta}-\bm{\theta}^* \right\rangle$. As $\mathbb{E}\left[  q(\bm{\theta}, X) \right] = \bm{\theta}-\bm{\theta}^*$, so
$
 \left\langle q(\bm{\theta}, X) - \mathbb{E}\left[  q(\bm{\theta}, X) \right], \mathbf{v} \right\rangle  
= \left\langle \bm{u}, \bm{\theta}-\bm{\theta}^* \right\rangle     \left\langle  \bm{u},  \mathbf{v}  \right\rangle - \left\langle \bm{\theta}-\bm{\theta}^*, \mathbf{v} \right\rangle.
$

For a fixed $\bm{\theta} \in \bm{\Theta}$, $\bm{\theta} \ne \bm{\theta}^*$, we let $\eth = \left\| \bm{\theta}-\bm{\theta}^*   \right\| >0 $. We further decompose $\bm{\theta}-\bm{\theta}^*$ as $\bm{\theta}-\bm{\theta}^* = \sqrt{c_1} \mathbf{v} + \sqrt{c_2} \mathbf{\hat{\mathbf{v}}}$, where $\mathbf{\hat{\mathbf{v}}}$ is an vector perpendicular to $\mathbf{v}$, $c_1 + c_2 = \eth^2$. We further have that $\left\langle \bm{u}, \mathbf{\hat{\mathbf{v}}}  \right\rangle \sim N(0,1)$ and: 
\begin{align}
\label{assum3_check_sec_equ}
&\left\langle \bm{u}, \bm{\theta}-\bm{\theta}^* \right\rangle  \left\langle  \bm{u},  \mathbf{v}  \right\rangle - \left\langle \bm{\theta}-\bm{\theta}^*, \mathbf{v} \right\rangle   \nonumber \\
&= \sqrt{c_1} \left( \left\langle \bm{u}, \mathbf{{\mathbf{v}}}  \right\rangle^2 -1  \right) + \sqrt{c_2} \left\langle \bm{u}, \mathbf{\hat{\mathbf{v}}}  \right\rangle  \left\langle \bm{u}, \mathbf{{\mathbf{v}}}  \right\rangle.
\end{align}

One also has
$
\mathbb{E}\left[  \left\langle \bm{u}, \mathbf{\hat{\mathbf{v}}}  \right\rangle \left\langle \bm{u}, \mathbf{{\mathbf{v}}}  \right\rangle   \right] 
= \mathbb{E}\left[ \mathbf{\hat{\mathbf{v}}}^\top \mathbf{u} \mathbf{u}^\top \mathbf{v}     \right] 
= \mathbf{\hat{\mathbf{v}}}^\top \mathbb{E}\left[  \mathbf{u} \mathbf{u}^\top  \right] \mathbf{v} = 0,
$
where $ \mathbf{u}^\top$ is the transpose of $\mathbf{u}$. Hence, $\left\langle \bm{u}, \mathbf{\hat{\mathbf{v}}}  \right\rangle$ and $\left\langle \bm{u}, \mathbf{{\mathbf{v}}}  \right\rangle$ are mutually independent.
For any $\varphi$ satisfies $\varphi \sqrt{c_1} <1/4$ and $\varphi^2 c_2 <1/4$, we have:
\begin{align}
&\mathbb{E}\left[ \text{exp}  \left(\varphi  \left\langle q(\bm{\theta}, X) - \mathbb{E}\left[  q(\bm{\theta}, X) \right], \mathbf{v} \right\rangle   \right)  \right] \nonumber \\
&\stackrel{(a)}= \mathbb{E}\left[  \text{exp}  \left(  \varphi \sqrt{c_1} \left( \left\langle \bm{u}, \mathbf{{\mathbf{v}}}  \right\rangle^2 -1  \right) + \varphi \sqrt{c_2} \left\langle \bm{u}, \mathbf{\hat{\mathbf{v}}}  \right\rangle  \left\langle \bm{u}, \mathbf{{\mathbf{v}}}  \right\rangle   \right)   \right] \nonumber \\
&\stackrel{(b)} \le \sqrt{\mathbb{E}\left[  e^{2\varphi \sqrt{c_1} \left( \left\langle \bm{u}, \mathbf{{\mathbf{v}}}  \right\rangle^2 -1  \right)} \right]  
	\mathbb{E}\left[ e^{2\varphi  \sqrt{c_2} \left\langle \bm{u}, \mathbf{\hat{\mathbf{v}}}  \right\rangle    \left\langle \bm{u}, \mathbf{{\mathbf{v}}}  \right\rangle  }   \right]     } \nonumber \\
&= e^{-\varphi \sqrt{c_1}} \sqrt{\mathbb{E}\left[  e^{2\varphi \sqrt{c_1} \left( \left\langle \bm{u}, \mathbf{{\mathbf{v}}}  \right\rangle^2 \right)} \right]  
	\mathbb{E}\left[ e^{2\varphi  \sqrt{c_2} \left\langle \bm{u}, \mathbf{\hat{\mathbf{v}}}  \right\rangle    \left\langle \bm{u}, \mathbf{{\mathbf{v}}}  \right\rangle  }   \right]     } \nonumber \\
&\stackrel{(c)}= e^{-\varphi \sqrt{c_1} } \left( 1 - 4\varphi \sqrt{c_1}  \right) ^{-1/4} \left( 1-4 \varphi^2 c_2   \right) ^{-1/4},
\end{align}
where $(a)$ holds by plugging in Eq.~(\ref{assum3_check_sec_equ}); $(b)$ is true by applying the Cauchy-Schwartz's inequality; $(c)$ is true by applying the moment generating function of $\chi^2$ distribution.
Since $1-t \ge e^{-4t}$ for $0 \le t \le 1/2$, and $e^{-t} / \sqrt{1-2t} \le e^{2t^2}$ for $|t| \le 1/4$. Thus, for $\varphi^2 \le 1/(64\eth^2)$, one has 
$
\mathbb{E}\left[ \text{exp}  \left(\varphi  \left\langle q(\bm{\theta}, X) - \mathbb{E}\left[  q(\bm{\theta}, X) \right], \mathbf{v} \right\rangle   \right)  \right]
 \le \text{exp} \left(  4 \varphi^2 (c_1 + c_2)  \right)  
\le \text{exp} \left(  4 \varphi^2 \eth^2  \right).
$
Therefore, Assumption~\ref{assumption_3} holds with $\alpha_2 = \sqrt{8}$ and $\rho_2 = 8$.

Last, we check Assumption~\ref{assumption_4}.
As $\nabla f(\bm{\theta}, X) = \bm{u} \left\langle \bm{u}, \bm{\theta}-\bm{\theta}^* \right\rangle -\bm{u} e$, then $\nabla^2 f(\bm{\theta}, X) = \bm{u} \bm{u}^{\top}$, thus it suffices to show the following
$
\mathbb{P} \left\{ {\left\| \frac{1}{\left| {X_s} \right|}  \sum\nolimits_{x \in {X_s}}   \nabla^2 f(\bm{\theta}, x) \right\|} 
\le  H  \right\} 
= \mathbb{P} \left\{ {\left\| \frac{1}{\left| {X_s} \right|}  \sum\nolimits_{j =1}^{\left| {X_s} \right|}  \bm{u}_j \bm{u}_j^{\top} \right\|} 
\le  H  \right\}
\ge 1-\beta/3.
$
Let $\bm{U}= \left[\bm{u}_1, \bm{u}_2,...,\bm{u}_{\left| {X_s} \right|} \right] \subset \mathbb{R}^{d \times {\left| {X_s} \right|}}$, one has $\sum\nolimits_{j =1}^{\left| {X_s} \right|}  \bm{u}_j \bm{u}_j^{\top} = \bm{U} \bm{U}^{\top} $ and 
$
\mathbb{P} \left\{ {\left\| \frac{1}{\left| {X_s} \right|}  \sum\nolimits_{j =1}^{\left| {X_s} \right|}   \bm{u}_j \bm{u}_j^{\top} \right\|} 
\le  H  \right\}  = \mathbb{P} \left\{ \left\| \bm{U}  \right\| \le \sqrt{\left| {X_s} \right| H} \right\}.
$
Since $\bm{U}$ is an i.i.d. standard Gaussian matrix, then according to~\cite{vershynin2010introduction}, for $t\ge 0$, we have that:
$
\mathbb{P} \left\{ \left\| \bm{U}  \right\| \le \sqrt{\left| {X_s} \right| } + \sqrt{d} +t \right\} \ge 1 - \text{exp} \left( -t^2/2  \right).
$
Setting $H = \left( \sqrt{\left| {X_s} \right|  } +\sqrt{d} +  \sqrt{2 \text{log} (4/\beta)} \right)^2 / {\left| {X_s} \right|}$ and  $t = \sqrt{2 \text{log} (4/\beta)}$ to complete the proof.

\subsection{Proof of Theorem~\ref{theorem_2}} \label{sec:appendix_2}

Since we assume that the server delay $\tau_s$ is zero, i.e., $\tau_s=0$. Then in the following, we use $\bm{g}_s^{t}$ to denote the server model update.

If server updates the global model following the AFLGuard algorithm, i.e., Algorithm~\ref{AFLGuard_lip_alg}, then for any $t > 0, \tau_i > 0  $, we have:
\begin{align}
\left\| \bm{\theta}^{t +1} - \bm{\theta}^* \right\| 
&= \left\| \bm{\theta}^{t} - \eta \bm{g}_i^{t- \tau_i} - \bm{\theta}^* \right\|   \nonumber\\
&\le \left\| \bm{\theta}^{t} - \eta \nabla F(\bm{\theta}^{t}) - \bm{\theta}^* \right\| + \eta \left\| \bm{g}_i^{t- \tau_i} - \nabla F(\bm{\theta}^{t}) \right\| \nonumber\\
& \stackrel{(a)} \le \underbrace{ \left\| \bm{\theta}^{t} - \eta \nabla F(\bm{\theta}^{t}) - \bm{\theta}^* \right\| }_{\clubsuit} +   \eta \lambda \underbrace{\left\| \nabla F(\bm{\theta}^{t}) - \nabla F(\bm{\theta}^*) \right\| }_{\bigstar} \nonumber\\
& \quad +   \eta( \lambda +1) \underbrace{\left\| \bm{g}_s^{t} - \nabla F(\bm{\theta}^{t}) \right\|}_{\blacklozenge} \nonumber\\
& \stackrel{(b)} \le \left( \sqrt{1 - \frac{2\eta \mu L }{\mu +L} }  + \eta L\lambda + 8\eta \Lambda ( \lambda +1)   \right)  \left\|  \bm{\theta}^{t} - \bm{\theta}^*  \right\| \nonumber \\ 
& \quad + 4\eta \Gamma  ( \lambda +1),
\end{align}
where $(a)$ uses $\nabla F(\bm{\theta}^*)=0$ and Lemma~\ref{lemma_4}, $(b)$ is true by plugging in Lemma~\ref{lemma_2}, Assumption~\ref{assumption_1} and Lemma~\ref{lemma_3} into $\clubsuit$, $\bigstar$, and $\blacklozenge$, respectively. 
Telescoping, one has
$
\left\| \bm{\theta}^t - \bm{\theta}^* \right\| 
\le
\left( 1- q \right)^t \left\| \bm{\theta}^0 - \bm{\theta}^* \right\| +  4\eta \Gamma ( \lambda +1) / q,
$
where $q = 1 -  \left( \sqrt{1 - {2\eta \mu L }/(\mu +L) }  + \eta L\lambda + 8\eta \Lambda ( \lambda +1)   \right) $.

Next, we proof Lemma~\ref{lemma_4}, Lemma~\ref{lemma_2} and Lemma~\ref{lemma_3} one by one.

\begin{lem}
	\label{lemma_4}
	If the server uses the AFLGuard algorithm to update the global model, then 
	for arbitrary number of malicious clients, one has:
	\begin{align}
	\left\| \bm{g}_i^{t-\tau_i} -  \nabla F(\bm{\theta}^{t}) \right\|  
	\le 
	(\lambda +1) \left\|  \bm{g}_s^{t}  - \nabla F(\bm{\theta}^{t}) \right\| +  \lambda \left\| \nabla F(\bm{\theta}^{t}) \right\|. \nonumber
	\end{align}
\end{lem}
\begin{proof}
	\begin{align}
& \left\| \bm{g}_i^{t-\tau_i}  - \nabla F(\bm{\theta}^{t}) \right\|  \nonumber\\
&\le \left\|  \bm{g}_i^{t-\tau_i} - \bm{g}_s^{t}  \right\|+  \left\| \bm{g}_s^{t} -  \nabla F(\bm{\theta}^{t}) \right\|  \nonumber\\
&\stackrel{(a)} \le  \lambda \left\|  \bm{g}_s^{t} \right\| + \left\| \bm{g}_s^{t} -  \nabla F(\bm{\theta}^{t}) \right\|  \nonumber\\
&\le \lambda \left\| \bm{g}_s^{t}  -\nabla F(\bm{\theta}^{t}) \right\| + \lambda \left\| \nabla F(\bm{\theta}^{t}) \right\| + \left\| \bm{g}_s^{t}  -  \nabla F(\bm{\theta}^{t}) \right\|  \nonumber\\
&= (\lambda +1) \left\| \bm{g}_s^{t}  - \nabla F(\bm{\theta}^{t})\right\| +  \lambda \left\| \nabla F(\bm{\theta}^{t}) \right\|  , 
\end{align}
	where $(a)$ is because for the AFLGuard algorithm, we have that $\left\|  \bm{g}_i^{t-\tau_i} - \bm{g}_s^{t}  \right\| \le \lambda \left\| \bm{g}_s^{t}  \right\|$.
\end{proof}

\begin{lem}
	\label{lemma_2}
	Suppose Assumption~\ref{assumption_1} holds, if the global learning rate satisfies $\eta \le \frac{2}{\mu+ L}$, then we have the following:
	\begin{align}
	\left\| \bm{\theta}^{t} - \eta \nabla F(\bm{\theta}^{t}) - \bm{\theta} ^* \right\| 
	\le \sqrt{1 - {2\eta \mu L }/(\mu +L) }  \left\| \bm{\theta}^ {t}  - \bm{\theta} ^* \right\|. \nonumber
	\end{align}
	\end{lem}
\begin{proof}
	$
	\left\| \bm{\theta}^{t} - \eta \nabla F(\bm{\theta}^{t}) - \bm{\theta} ^*  \right\|^2 
	 = \left\| \bm{\theta}^{t} - \bm{\theta} ^* \right\|^2 + \eta^2 \left\| \nabla F(\bm{\theta}^{t})  \right\|^2 
	 -2 \eta \left\langle { \bm{\theta}^{t}-\bm{\theta} ^*, F(\bm{\theta}^{t}) }\right\rangle .
	\label{one_step_descent_square}
	$
	According to~\cite{bubeck2014convex}, if $F(\bm{\theta})$ is $L$-smooth and
	$\mu$-strongly convex, for any $\bm{\theta}, \bm{\theta}^{\prime} \in \Theta$, one has
		$
	\label{convex_smooth_lemma}
	\frac{\mu L}{\mu + L} \left\| \bm{\theta} - \bm{\theta}^{\prime} \right\|^2 + \frac{1}{\mu + L} \left\| \nabla F(\bm{\theta}) - \nabla F(\bm{\theta}^{\prime}) \right\|^2 
	 \le 
	\left\langle { \nabla F(\bm{\theta}) - \nabla F(\bm{\theta}^{\prime})},\bm{\theta} - \bm{\theta}^{\prime} \right\rangle.
	$
	Setting $\bm{\theta} = \bm{\theta}^{t}, \bm{\theta}^{\prime} =\bm{\theta} ^*$, since $\nabla F(\bm{\theta}^*)=0$, we have that
		$
	\label{convex_smooth_lemma_one_step}
	\frac{\mu L}{\mu + L} \left\| \bm{\theta}^{t} - \bm{\theta} ^* \right\|^2 + \frac{1}{\mu + L} \left\| \nabla F(\bm{\theta}^{t}) \right\|^2 
	\le 
	\left\langle  \nabla F(\bm{\theta}^{t}),\bm{\theta}^{t} - \bm{\theta} ^*  \right\rangle.
		$
	Thus one has:
	\begin{align} 
	& \left\| \bm{\theta}^{t} - \eta \nabla F(\bm{\theta}^{t}) - \bm{\theta} ^*  \right\|^2 
	\le \left\| \bm{\theta}^{t} - \bm{\theta} ^* \right\|^2 + \eta^2 \left\| \nabla F(\bm{\theta}^{t})  \right\|^2 \nonumber\\
	&\quad -2 \eta \left( \frac{\mu L}{\mu + L} \left\| \bm{\theta}^{t}  - \bm{\theta}^* \right\|^2  + \frac{1}{\mu +L}  \left\| \nabla F(\bm{\theta}^{t})  \right\|^2  \right) \nonumber\\
	&= \left(  1 - {2\eta \mu L }/(\mu +L)\right)  \left\| \bm{\theta}^{t} - \bm{\theta} ^* \right\|^2 + \eta \left(  \eta - {2}/(\mu +L)\right) \left\| \nabla F(\bm{\theta}^{t})  \right\|^2 \nonumber\\
	&\stackrel{(a)} \le \left(  1 - {2\eta \mu L }/(\mu +L)\right) \left\| \bm{\theta}^{t} - \bm{\theta} ^* \right\|^2,
	\label{last_Equ_first_part}
	\end{align}
	where $(a)$ is because $0 \! < \! \eta \! \le \! {2}/(\mu +L)$.
	$
	\label{square_lemma}
	\left\| \bm{\theta}^{t} - \eta \nabla F(\bm{\theta}^{t}) - \bm{\theta} ^*  \right\|
	\! \le  \!
	\sqrt{1 - {2\eta \mu L }/ (\mu +L) } 	\left\|  \bm{\theta}^{t} - \bm{\theta} ^* \right\|. 
	$
\end{proof}	
The proof of Lemma~\ref{lemma_3} is mainly motivated from~\cite{ChenPOMACS17,cao2020fltrust}.
To simplify the notation, we will ignore the superscript $t$ in $\bm{g}_s^t$. 
Define 	$\nabla \bar{f_s}(\bm{\theta}) = \frac{1}{\left| {X_s} \right|}\sum\nolimits_{x \in {X_s}} \nabla f(\bm{\theta},x)$.

\begin{lem}
	\label{lemma_3}
	If Assumptions~\ref{assumption_2}-\ref{assumption_4} hold and  
	$\Theta  \subset \{ {\bm{\theta}:\left\| \bm{\theta} - {\bm{\theta}^*} \right\| \le \epsilon \sqrt d } \}$ holds for some parameter $\epsilon > 0$.   For any  $\beta  \in (0,1)$, if $\Gamma  \le \alpha_1 ^2 / \rho_1$ and $\Lambda  \le \alpha_2 ^2 / \rho_2$, we have that:
	\begin{align}
	\mathbb{P} \left\{ {\left\| \bm{g}_s - \nabla F(\bm{\theta}) \right\| \le 8\Lambda {\left\| {\bm{\theta} - \bm{\theta}^*} \right\|} + 4\Gamma } \right\} \ge 1 - \beta,  \nonumber
	\end{align}
%
	where $\Gamma, \Lambda$ are defined in Theorem~\ref{theorem_2}.
\end{lem}
\begin{proof}
	We define $\xi = \frac{\rho_2\alpha_1}{2\alpha_2 ^2}\sqrt{\frac{d}{\left| X_s \right|}}$ and let $\psi =\left\lceil \epsilon \sqrt d/\xi \right\rceil $.
	Then for any integer $1 \le l \le \psi$, we define 
	$
	{\Theta_l } = \left\{ {\bm{\theta}:\left\| {\bm{\theta} - {\bm{\theta}^*}} \right\| \le \xi l } \right\}.
	$
	For a given integer $l$, we let $\bm{\theta}_1,\cdots,\bm{\theta}_{\hat{\varsigma}}$ be an $\omega$-cover of ${\Theta _l }$, where $\omega = \frac{{\alpha_2}\xi l }{R}\sqrt {d/{\left| {X_s} \right|}}  $, where ${R} = \max \left\{ {L,{H}} \right\}$.
	 From~\cite{vershynin2010introduction}, we know that $\log \hat{\varsigma} \le d\log ({3\xi l / \omega}) $. 
	 For any $\bm{\theta} \in {\Theta_l}$, there exists a $1 \le c \le \omega $ such that $\left\| {\bm{\theta} - \bm{\theta}_{c} } \right\| \le \omega$ holds.
	Then, based on the triangle inequality, one has
	$
	 \left\| \nabla \bar{f_s}(\bm{\theta}) - \nabla F(\bm{\theta}) \right\| 
	\le 
	 \left\| {\nabla F(\bm{\theta}) - \nabla F({\bm{\theta}_{c} })} \right\|  
	 + \left\| \nabla \bar{f_s}(\bm{\theta}) - \nabla \bar{f_s}(\bm{\theta}_c)   \right\| 
	  + \left\| \nabla \bar{f_s}(\bm{\theta}_c)  - \nabla F({\bm{\theta}_{c} })   \right\| .
	$
	By Assumption~\ref{assumption_1}, one has
	$
	\label{T_1_equ}
	\left\| {\nabla F(\bm{\theta}) - \nabla F({\bm{\theta}_{c}})} \right\| \le L\left\| {\bm{\theta} - \bm{\theta}_{c}} \right\| \le L \omega.
	$
	We define event $E_1$ as:
	\begin{align}
	\!\!\! E_1 \!=\! \left\{ \mathop {\sup }\nolimits_{\bm{\theta},\bm{\theta}^{\prime} \in \Theta :\bm{\theta} \ne \bm{\theta}^{\prime}}
	 \left\| \nabla \bar{f_s}(\bm{\theta}) \!-\! \nabla \bar{f_s}(\bm{\theta}^{\prime}) \right\| \!\le\! H \left\| \bm{\theta} \!-\! \bm{\theta}^{\prime} \right\|  \right\}.
	\end{align}

	According to Assumption~\ref{assumption_4}, we have $\mathbb{P} \left\{ E_1 \right\} \ge 1 - {\beta }/{3}$. One also has
    $
	\label{T_2_equ}
	 \mathop {\sup }\nolimits_{\bm{\theta} \in \Theta } \left\| \nabla \bar{f_s}(\bm{\theta}) - \nabla \bar{f_s}(\bm{\theta}_c) \right\| 
	 \le  H \omega.
    $
	By the triangle inequality again, and because $\mathbb{E}\left[ {q\left( \bm{\theta}, X \right)} \right] = \nabla F(\bm{\theta}) - \nabla F({\bm{\theta}^*})$, we have:
	\begin{equation}
	\begin{aligned}[b]
	& \left\| \nabla \bar{f_s}(\bm{\theta}_c)  - \nabla F(\bm{\theta}_c)    \right\|  
	 \le  \left\| \nabla \bar{f_s}(\bm{\theta}^{*}) - \nabla F({\bm{\theta}^*}) \right\|   \\
	&  \quad +  \left\| \nabla \bar{f_s}(\bm{\theta}_c) -  \nabla \bar{f_s}(\bm{\theta}^*)  -  \left(  \nabla F(\bm{\theta}_c)
	- \nabla F(\bm{\theta}^*)
	 \right)     \right\| 	 \\
	&   \le  \left\|  \nabla \bar{f_s}(\bm{\theta}^*) - \nabla F(\bm{\theta}^*) \right\|   
	+
	 \left\| {\frac{1}{{\left| {{X_s}} \right|}}  \sum\nolimits_{x \in {X_s}} {q(\bm{\theta}_{c}, {x}) - \mathbb{E} \left[ {q\left( {\bm{\theta}_{c}}, X\right)} \right]} } \right\|. \nonumber
	\end{aligned}
	\end{equation}
	
	Define events $E_2 = \left\{  \left\| \nabla \bar{f_s}(\bm{\theta}^*) -\nabla F(\bm{\theta}^*) \right\| \le 2{\Gamma}  \right\}$ and $E_l$ as:
	\begin{equation}
	\begin{aligned}[b]
	 E_l = \left\{  \mathop {\sup }\nolimits_{1 \le k \le {N}}  {\left\| \frac{1}{{\left| {X_s} \right|}}\sum\nolimits_{x \in {X_s}} {q(\bm{\theta}_k, x) - \mathbb{E} \left[ {q\left( \bm{\theta}_k, X \right)} \right]}  \right\|  }  
	\right.  
	 \left.	\le 2 \Lambda \xi l \vphantom { \mathop {\sup }\limits_{1 \le {j } \le {\left| D \right|}_{\varepsilon}} }  \right\}. \nonumber 
	\end{aligned}
	\end{equation}

	 By Proposition~\ref{proposition_1}, Proposition~\ref{proposition_2}, since $\Gamma \le \alpha_1 ^2 / \rho_1$, $\Lambda \le \alpha_2 ^2 / \rho_2$, we have $\mathbb{P} \left\{ E_2 \right\} \ge 1 - {\beta }/{3}$, $\mathbb{P} \left\{ E_l \right\} \ge 1 - {\beta }/({3\psi})$.
	Thus, on event $E_1 \cap  E_2 \cap  E_l$, one has
    $
	\mathop {\sup }\nolimits_{\bm{\theta} \in {\Theta _l }} \left\| \nabla \bar{f_s}(\bm{\theta}) - \nabla F(\bm{\theta})  \right\| 
	\le L \omega + H \omega  + 2{\Gamma} + 2{\Lambda} \xi l 	
	\le 4{\Lambda}\xi l + 2{\Gamma}.
	$
	Thus, we have at least $1 - \beta$ that event  $ E = E_1 \cap  E_2 \cap (\cap_{l=1}^{ \psi } E_l )$.
	Also, on event $  E$, for any $\bm{\theta} \in \Theta _{\psi}$, there exists an $1 \le l \le \psi$ such that $(l -1)\xi  < \left\| {\bm{\theta} - \bm{\theta}^*} \right\| \le \xi l$ holds.
	If $l =1$, we have
	$
	\left\| \nabla \bar{f_s}(\bm{\theta}) - \nabla F(\bm{\theta}) \right\|  
	 \le 4{\Lambda} \xi + 2{\Gamma}
	 \le 4 \Gamma;
	$
	and $2(l -1) \ge l$ if $l \ge 2$.
	Thus, one has
	$
	\label{ell_2_bound}
	\left\| \nabla \bar{f_s}(\bm{\theta}) - \nabla F(\bm{\theta}) \right\|  
	\le 8{\Lambda} \left\| {\bm{\theta} - \bm{\theta}^*} \right\| + 2{\Gamma }.
	$
	On $E$, one has
	$
	\mathop {\sup }\nolimits_{\bm{\theta} \in {\Theta _{\psi} }} \left\| \nabla \bar{f_s}(\bm{\theta}) - \nabla F(\bm{\theta}) \right\|  
	 \le 8{\Lambda} \left\| {\bm{\theta} - \bm{\theta}^*} \right\| + 4{\Gamma }.
	$
\end{proof}

The following proof of Proposition~\ref{proposition_1} is mainly motivated from~\cite{ChenPOMACS17,cao2020fltrust}.
\begin{proposition}
	\label{proposition_1}
	Suppose Assumption~\ref{assumption_2} holds. For any $\beta  \in (0,1)$, $\bm{\theta} \in \Theta$, let ${\Gamma } = \sqrt 2 {\alpha_1 } \sqrt {(d\log 6 + \log (3/ \beta))/ {\left| {X_s} \right|}}$.
	If ${\Gamma } \le {\alpha_1^2} / {\rho_1}$, then we have:
	\begin{align}
	& \mathbb{P} \left\{ {\left\| {\frac{1}{\left| X_s \right|}\sum\nolimits_{x \in {X}} {\nabla f(\bm{\theta}^*, x)}  - \nabla F({\bm{\theta}^*})} \right\| \ge 2{\Gamma }} \right\} \le {\beta }/{3}. \nonumber
	\end{align}
\end{proposition}

\begin{proof}
	 We let
	 $\mathbf{B} = \{ {\mathbf{v}_{1,}}, \cdots, {\mathbf{v}_{\varsigma }\}}$ be one $\frac{1}{2}$-cover of the unit sphere $\mathbf{V}$.
	By ~\cite{vershynin2010introduction}, one has $\log \varsigma \le d\log 6$ and
	$
	\label{vershynin2010introduction_equ}
	\left\| \nabla \bar{f_s}(\bm{\bm{\theta}^*})  - \nabla F({\bm{\theta}^*}) \right\|  
	\le 
	2\mathop {\sup }\nolimits_{\mathbf{v} \in \mathbf{B}} \left\{ {\left\langle {\nabla \bar{f_s}(\bm{\theta}^*)  - \nabla F(\bm{\theta}^*),\mathbf{v}} \right\rangle } \right\}. 
	$
	If Assumption~\ref{assumption_2} and the condition $ \Gamma  \le \alpha_1 ^2 / \rho_1$ satisfy, and according to the concentration inequalities for sub-exponential random variables~\cite{wainwright2019high}, we have that
	$
	 \mathbb{P} \left\{ {\left\langle {\nabla \bar{f_s}({\bm{\theta}^*})  - \nabla   F({\bm{\theta}^*}),\mathbf{v}} \right\rangle  	\ge {\Gamma }} \right\} 
	 \le \exp \left( { - \left| {{X_s}} \right|\Gamma ^2} / (2 \alpha_1  ^2) \right).
	$
	One has
	$
	\label{last_equ_lemma3}
	  \mathbb{P} \left\{ {\left\| {\nabla \bar{f_s}({\bm{\theta}^*}) - \nabla F({\bm{\theta}^*})} \right\| \ge 2{\Gamma }} \right\} 
	\le  \exp \! \left( { - \left| {{X_s}} \right|\Gamma ^2} / (2\alpha_1 ^2) + d\log 6 \right)
	$
	by the union bound.
	Put in ${\Gamma }$ finishes the proof.
\end{proof}

\begin{proposition}
	\label{proposition_2}
	Suppose Assumption~\ref{assumption_3} holds. For any $\beta  \in (0,1)$, $\bm{\theta} \in \Theta$, let ${\Delta} = \sqrt 2 {\alpha_2} \sqrt {(d\log 6 + \log (3 / \beta))/ {\left| {X_s} \right|}}$. 
	If ${\Delta} \le {\alpha_2^2} / {\rho_2}$, then we have:
	\begin{align}
	\mathbb{P} \left\{ \left\| {\frac{1}{{\left| {{X_s}} \right|}}\sum\limits_{x \in {X_s}} {\nabla q(\bm{\theta},x)}  - {\mathbb{E}\left[ {q(\bm{\theta},X)} \right]}} \right\| 
	\ge 2{\Delta} {\left\| {\bm{\theta} - {\bm{\theta}^*}} \right\|}  \vphantom{\frac{1}{\left| {D_0} \right|} \sum\limits_{X_i \in {D_0}}}  \right\}  
	\le {\beta }/{3}.  \nonumber
	\end{align}
	%
\end{proposition}
\begin{proof}
	The proof of Proposition~\ref{proposition_2} is similar to that of Proposition~\ref{proposition_1}, and is omitted here for brevity.
\end{proof}

\begin{figure*}[!t]
	\centering
	\includegraphics[scale = 0.5]{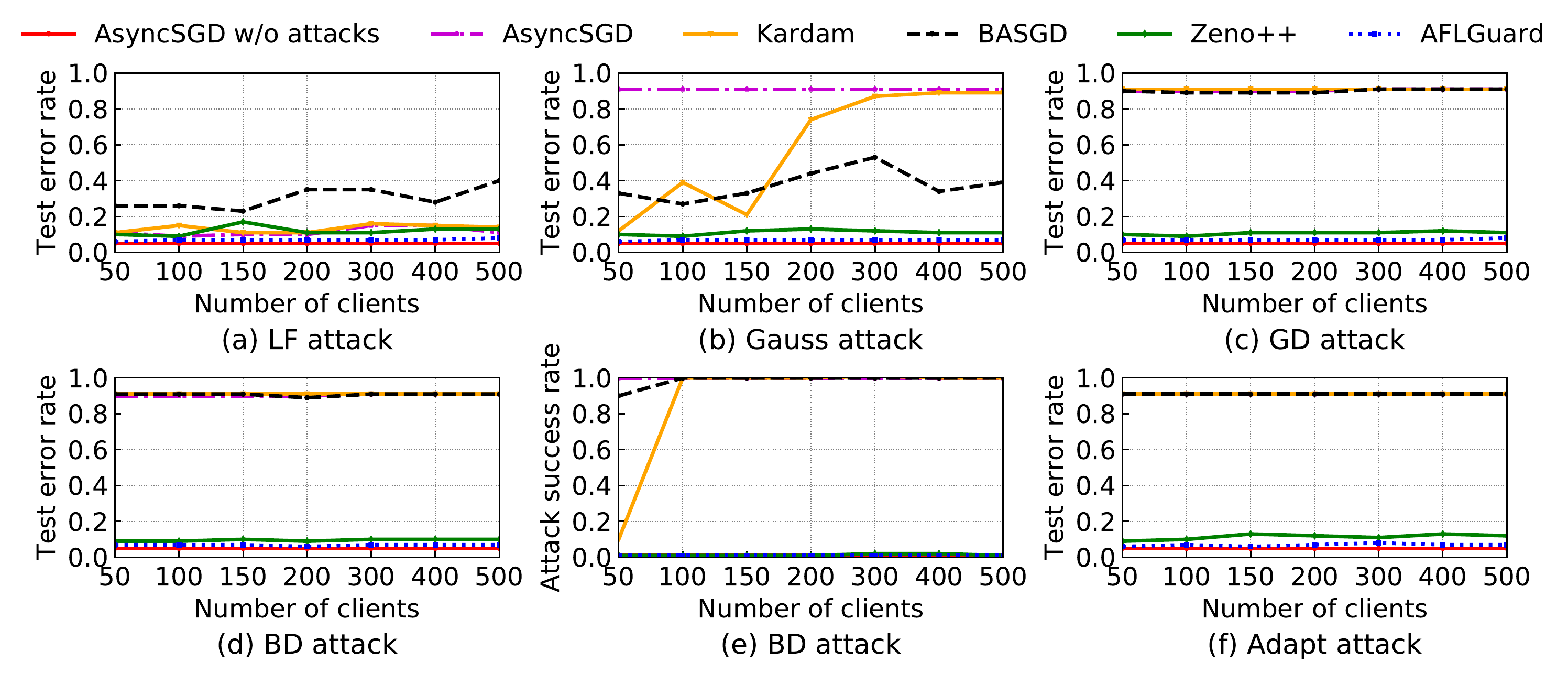}
	\caption{Test error rates and attack success rates of different defenses under different attacks with different number of clients on MNIST dataset.}
	\label{attack_client_number}
\end{figure*}

\subsection{Datasets} \label{sec:appendix_datasets}

\myparatight{1) Synthetic Dataset} 
We randomly generate 10,000 data samples of dimensions $d=100$.
Each dimension follows the Gaussian distribution $N(0,1)$ and noise $e_i$ is sampled from $N(0,1)$.
We use $N(0, 25)$ to generate each entry of $\bm{\theta}^*$.
We generate $y_i$ according to the linear regression model in Lemma~\ref{lemma_assum}.
We randomly draw 8,000 samples for training and use the remaining 2,000 samples for testing.

\myparatight{2) MNIST~\cite{lecun2010mnist}} MNIST is a 10-class handwritten digits image classification dataset, which contains 60,000 samples for training and 10,000 examples for testing. 

\myparatight{3) Fashion-MNIST~\cite{xiao2017online}} Fashion-MNIST is a dataset containing images of 70,000 fashion products from 10 classes. The training set has 60,000 images and the testing set has 10,000 images.

\myparatight{4) Human Activity Recognition (HAR) ~\cite{anguita2013public}} The HAR dataset aims to recognize 6 types of human activities 
and the dataset is collected from smartphones of 30 real-world users. There are 10,299 examples in total and each example includes 561 features.
We randomly sample 75\% of each client's examples as training data and use the rest as test data.

\myparatight{5) Colorectal Histology MNIST~\cite{kather2016multi}} Colorectal Histology MNIST is an 8-class dataset for classification of textures in human colorectal cancer histology. This dataset contains 5,000 images and each image has 64×64 grayscale pixels. We randomly select 4,000 images for training and use the remaining 1,000 images for testing.

\myparatight{6) CIFAR-10~\cite{krizhevsky2009learning}} CIFAR-10 consists of 60,000 color images. 
This dataset has 10 classes, and there are 6,000 images of each class. 
The training set has 50,000 images and the testing set has 10,000 images.

\begin{table}[htb!]
	\caption{The CNN architecture.}
	\centering
	  	\vspace{1mm}
	\small
	\begin{tabular}{|c|c|} \hline 
		{Layer} & {Size} \\ \hline
		{Input} & { $28\times28\times1$}\\ \hline
		{Convolution + ReLU} & { $3\times3\times30$}\\ \hline
		{Max Pooling} & { $2\times2$}\\ \hline
		{Convolution + ReLU} & { $3\times3\times50$}\\ \hline
		{Max Pooling} & { $2\times2$}\\ \hline
		{Fully Connected + ReLU} & {100}\\ \hline
		{Softmax} & {10}\\ \hline
	\end{tabular}
	\label{cnn_arch}
\end{table}

\begin{table*}[htbp]
	\centering
\caption{Test error rates and attack success rates of Zeno++ and AFLGuard under different attacks with different distribution shifts (DSs) on Fashion-MNIST, HAR, Colorectal Histology MNIST and CIFAR-10 datasets. The results of BD attack are in the form of “test error rate / attack success rate”.}
	\vspace{1mm}
	\label{bias_single_app}%
		\small 
\begin{tabular}{c}
\textbf{\footnotesize (a) Fashion-MNIST} \\
\end{tabular}\\
	{
		\begin{tabular}{|c|c|c|c|c|c|c|c|c|c|c|}
			\hline
			DS   & \multicolumn{2}{c|}{0.1} & \multicolumn{2}{c|}{0.5} & \multicolumn{2}{c|}{0.6} & \multicolumn{2}{c|}{0.8} & \multicolumn{2}{c|}{1.0} \\
			\hline
			Attack 
			&\tabincell{c}{Zeno++}   & \tabincell{c}{AFLGuard}   & \tabincell{c}{Zeno++}   & \tabincell{c}{AFLGuard}   & \tabincell{c}{Zeno++}   & \tabincell{c}{AFLGuard}   &\tabincell{c}{Zeno++}   & \tabincell{c}{AFLGuard}   & \tabincell{c}{Zeno++}   & \tabincell{c}{AFLGuard} \\
			\hline
			No attack  & 0.25 & 0.16 &  0.26  &  0.17 &  0.31 & 0.18 & 0.58  & 0.21 & 0.90  &  0.22 \\
			\hline
			LF attack  & 0.25 & 0.18  &    0.29   &  0.21     &  0.32      &   0.21    & 0.72      &   0.22     &  0.90     & 0.22 \\
			\hline
			Gauss attack   & 0.26  & 0.18 &  0.28     &   0.19  &  0.34     & 0.19      &    0.58   &   0.22    &   0.90    &0.23  \\
			\hline
			GD attack &  0.26 & 0.18 &   0.29     &   0.21      & 0.35      &  0.21     &    0.58   &  0.21   &   0.90    & 0.23 \\
			\hline
			BD attack  & 0.26 / 0.04 & 0.17 / 0.04  & 0.29 / 0.05   & 0.20 / 0.04  & 0.33 / 0.04  & 0.20 / 0.04 & 0.58 / 0.03 & 0.20 / 0.03   & 0.90 / 0.01 & 0.22 / 0.02 \\
			\hline
			Adapt attack  & 0.26  & 0.19  &  0.29    & 0.21      & 0.36   & 0.21    & 0.72    & 0.22   &   0.90    & 0.25 \\
			\hline
		\end{tabular}%
	}
	\\
	\vspace{0.05in}
\begin{tabular}{c}
\textbf{\footnotesize (b) HAR} \\
\end{tabular}\\
	{
		\begin{tabular}{|c|c|c|c|c|c|c|c|c|c|c|}
			\hline
			DS   & \multicolumn{2}{c|}{0.167} & \multicolumn{2}{c|}{0.5} & \multicolumn{2}{c|}{0.6} & \multicolumn{2}{c|}{0.8} & \multicolumn{2}{c|}{1.0} \\
			\hline
			Attack 
			&\tabincell{c}{Zeno++}   & \tabincell{c}{AFLGuard}   & \tabincell{c}{Zeno++}   & \tabincell{c}{AFLGuard}   & \tabincell{c}{Zeno++}   & \tabincell{c}{AFLGuard}   &\tabincell{c}{Zeno++}   & \tabincell{c}{AFLGuard}   & \tabincell{c}{Zeno++}   & \tabincell{c}{AFLGuard} \\
			\hline
			No attack  & 0.06 & 0.05  &  0.06    & 0.05      & 0.08   & 0.05    & 0.10    & 0.07   &   0.43    & 0.12 \\
			\hline
			LF attack & 0.07 & 0.05  &  0.08    & 0.05      & 0.09   & 0.05    & 0.10    & 0.09   &   0.43    & 0.36 \\
			\hline
			Gauss attack   & 0.07 & 0.05  &  0.07    & 0.05      & 0.09   & 0.05    & 0.10    & 0.07   &   0.43    & 0.36 \\
			\hline
			GD attack & 0.07 & 0.05  &  0.08    & 0.05     & 0.09   & 0.06    & 0.12    & 0.09  &   0.43    & 0.42 \\
			\hline
			BD attack  & 0.06 / 0.07 & 0.05 / 0.01  & 0.07 / 0.01   & 0.05 / 0.01  & 0.09 / 0.02 &  0.05 / 0.01  & 0.10 / 0.01 & 0.07 / 0.01  & 0.43 / 0.01 & 0.36 / 0.01 \\
			\hline
			Adapt attack  & 0.07 & 0.05  &  0.08    & 0.05      & 0.09   & 0.06    & 0.14     & 0.09   &   0.55    & 0.54 \\
			\hline
		\end{tabular}%
	}
	\\
		\vspace{0.05in}
\begin{tabular}{c}
\textbf{\footnotesize (c) Colorectal Histology MNIST} \\
\end{tabular}\\
	{
		\begin{tabular}{|c|c|c|c|c|c|c|c|c|c|c|}
			\hline
			DS   & \multicolumn{2}{c|}{0.125} & \multicolumn{2}{c|}{0.5} & \multicolumn{2}{c|}{0.6} & \multicolumn{2}{c|}{0.8} & \multicolumn{2}{c|}{1.0} \\
			\hline
			Attack 
			&\tabincell{c}{Zeno++}   & \tabincell{c}{AFLGuard}   & \tabincell{c}{Zeno++}   & \tabincell{c}{AFLGuard}   & \tabincell{c}{Zeno++}   & \tabincell{c}{AFLGuard}   &\tabincell{c}{Zeno++}   & \tabincell{c}{AFLGuard}   & \tabincell{c}{Zeno++}   & \tabincell{c}{AFLGuard} \\
			\hline
			No attack  & 0.25 & 0.18 &  0.31 & 0.22& 0.49 & 0.29  &0.62 & 0.32 &  0.71 & 0.34 \\
			\hline
			LF attack   & 0.31  & 0.21  &  0.39     &      0.23   &    0.53   &  0.37    &   0.63    &   0.41   &  0.88    &  0.41\\
			\hline
			Gauss attack   & 0.35  & 0.21  &  0.43     &    0.22   &  0.59     &    0.32   &   0.70    &   0.35   & 0.86      &  0.35 \\
			\hline
			GD attack  & 0.29 & 0.21   &   0.39    &  0.32     & 0.66      & 0.36      &   0.74    &  0.49    &    0.88   & 0.58 \\
			\hline
			BD attack  & 0.42 / 0.02 & 0.22 / 0.02  & 0.44 / 0.02   & 0.27 / 0.02   & 0.57 / 0.01 & 0.31 / 0.02 & 0.62 / 0.01 & 0.32 / 0.01   & 0.82 / 0.24 & 0.51 / 0.03 \\
	    	\hline
			Adapt attack  & 0.44 & 0.29  &  0.64    & 0.33      & 0.72   & 0.43    & 0.77    & 0.51   &   0.88    & 0.62 \\
			\hline
		\end{tabular}%
	}
	\\
		\vspace{0.05in}
\begin{tabular}{c}
\textbf{\footnotesize (d) CIFAR-10} \\
\end{tabular}\\
	{
		\begin{tabular}{|c|c|c|c|c|c|c|c|c|c|c|}
			\hline
			DS   & \multicolumn{2}{c|}{0.1} & \multicolumn{2}{c|}{0.5} & \multicolumn{2}{c|}{0.6} & \multicolumn{2}{c|}{0.8} & \multicolumn{2}{c|}{1.0} \\
			\hline
			Attack 
			&\tabincell{c}{Zeno++}   & \tabincell{c}{AFLGuard}   & \tabincell{c}{Zeno++}   & \tabincell{c}{AFLGuard}   & \tabincell{c}{Zeno++}   & \tabincell{c}{AFLGuard}   &\tabincell{c}{Zeno++}   & \tabincell{c}{AFLGuard}   & \tabincell{c}{Zeno++}   & \tabincell{c}{AFLGuard} \\
			\hline
			No attack & 0.32 & 0.24 & 0.41 & 0.26 & 0.54  & 0.29 &0.68 & 0.31  & 0.90  &  0.31\\
			\hline
			LF attack  & 0.33 &  0.34 &  0.53     &    0.34   &  0.64     &  0.35     &  0.71     &    0.35   & 0.90      & 0.38 \\
			\hline
			Gauss attack & 0.33 & 0.32  &  0.52     &   0.33    &  0.72     &  0.33     &   0.76    &   0.35    &    0.90   & 0.35 \\
			\hline
			GD attack & 0.32 & 0.27 &  0.60     &  0.30     &   0.83    &   0.31     &  0.85     &  0.33     &    0.90    & 0.32 \\
			\hline
			BD attack  & 0.32 / 0.95 & 0.28 / 0.01  & 0.49 / 0.06   & 0.29 / 0.01   & 0.62 / 0.00  &  0.32 / 0.02 & 0.80 / 0.01 &  0.34 / 0.01 & 0.90 / 0.00 & 0.36 / 0.04 \\
	 		\hline
			Adapt attack  & 0.77  & 0.32  &  0.82    & 0.36      & 0.90   & 0.36     & 0.90    & 0.36    &   0.90    & 0.39 \\
			\hline
		\end{tabular}%
	}
	\vspace{-0.25in}
\end{table*}%


\end{document}